%% file: paper.tex
\documentclass[prd,nofootinbib,twocolumn]{revtex4}
 \usepackage{graphicx}
\usepackage{latexsym} \usepackage{amsmath}
\usepackage{bm}
\usepackage{subfigure}

\def\bbh#1{binary black hole#1 (BBH#1)\gdef\bbh{BBH}}

\newcommand{\fmin}{{f_{\mathrm{min}}}}
\newcommand{\fmax}{{f_{\mathrm{max}}}}
\newcommand{\Mminimax}{{M_{\mathrm{minimax}}}}
\newcommand{\Mbest}{{M_{\mathrm{best}}}}
\newcommand{\Mtyp}{{M_{\mathrm{typ}}}}
\newcommand{\Omax}{{O_{\mathrm{max}}}}
\newcommand{\Hz}{{\mathrm{Hz}}}

\begin{document}

\title{Matched Filtering of Numerical Relativity Templates of Spinning Binary
Black Holes}
\author{Birjoo Vaishnav} 
\affiliation{Center for Gravitational Wave Physics, Penn State,
University Park, PA 16802}
\author{Ian Hinder}
\affiliation{Center for Gravitational Wave Physics, Penn State,
University Park, PA 16802}
\author{Frank Herrmann}
\affiliation{Center for Gravitational Wave Physics, Penn State,
University Park, PA 16802}
\author{Deirdre Shoemaker}
\affiliation{ Center for Gravitational Wave Physics,
Institute of Gravitational Physics and Geometry, Department of Physics,
Penn State, University Park, PA 16802}

\begin{abstract} 
Tremendous progress has been made towards the solution of the
binary-black-hole problem in numerical relativity.  The waveforms
produced by numerical relativity will play a role in gravitational
wave detection as either test-beds for analytic template banks 
or as template banks themselves.  As the parameter
space explored by numerical relativity expands, the importance of
quantifying the effect that each parameter has on 
first the detection of gravitational waves and then the parameter estimation 
of their sources increases.  In light of this, we present a study of
equal-mass, spinning binary-black-hole evolutions
through matched filtering techniques commonly used in data analysis. 
We study how the match between two numerical waveforms varies with 
numerical resolution, initial angular momentum of the black holes
and the inclination angle between the source and the detector.
This study is limited by the fact that the
spinning black-hole-binaries are oriented axially and the waveforms
only contain approximately two and a half orbits before merger.
We find that for detection purposes, 
spinning black holes require the inclusion of the higher harmonics
in addition to the dominant mode, a condition that becomes more important
as the black-hole-spins increase.
In addition, we conduct a preliminary investigation of how well a template 
of fixed spin and inclination angle can detect target templates of arbitrary spin
and inclination for the  axial case considered here.  
\end{abstract}

\maketitle

\section{Introduction}
\label{intro}
Initial LIGO has reached its design
sensitivity \cite{LIGO2006} and, along with other ground-based detectors
around the world, is taking science data.
The matched filtering technique employed to search for gravitational wave signals from
inspiraling and merging compact binaries is at
its best when accurate representations of the signal are used.  
One of the best sources for early detection is the inspiral and merger
of two black holes.   The accurate waveforms of the \bbh{} signal come from two sources:
post-Newtonian theory and numerical relativity.  Waveforms from post-Newtonian
theory are well understood during the inspiral phase of the binary evolution
up to an as yet unknown point close to the final plunge of the black holes.  
With numerical relativity, we attempt to 
determine the point at which the approximation breaks down and to supply the
waveform for the last orbit(s) and merger. Because black holes
in the $30-300M_\odot$ \cite{FHI} mass range are expected to merge at 
frequencies that are in the most sensitive part of the LIGO
frequency band,  the computation of \bbh{} waveforms has been a priority
for the numerical relativity community.
Fortunately, remarkable progress continues to be made in numerical relativity,
especially in solving the last orbits and merger of binary black hole systems.  

Since the breakthroughs in numerical relativity that allowed the
evolution of \bbh{s} through an orbit, merger and ringdown 
\cite{Pretorius:2005gq, Baker05a, Campanelli05a}, 
many groups have achieved similar success.
The numerical relativity has now begun evolving many orbits
\cite{baker-2006,buonanno-2006,pfeiffer-2007,ajith-2007} and exploring the parameter space
such as mass ratio and spins \cite{shoemaker2006,gonzalez-2007-98,CampanelliSpin,campanelli-2006-74,campanelli-2006,Herrmann07,koppitz-2007,gonzalez-2007,campanelli-2007-659,campanelli-2007,tichy-2007}. 
Each evolution of a \bbh{} system requires large amounts of computational resources.  The
resource requirements for longer evolutions with unequal masses and
spins are larger still.  Numerical relativity efforts will benefit
from {\em a priori} knowledge of where best to spend those resources to
create an appropriate library of waveforms.  Additionally, the cost of
doing data analysis for a large template bank with many parameters is
also computationally expensive.  Knowledge of the sensitivity of the
waveforms to the parameters will guide efforts in data analysis and numerical relativity.  As
the library of numerical relativity waveforms grows, we can determine
how the waveforms will be best employed in the search for
gravitational waves and the characterization of their sources.  They
may be used in conjunction with post-Newtonian waveforms through stitching, 
as templates for binary mergers, or to test the quality of template spaces built from
post-Newtonian and/or approximate waveforms.

In \cite{pan-2007}, a comparison of the closeness
between current gravitational wave search template banks and
numerical relativity waveforms from non-spinning \bbh{s} has
been performed.  The numerical relativity waveforms were used as 
target signals against which the template banks currently in use 
by LIGO to search for inspiral gravitational waves were tested. 
They found matches greater than 0.96 for many of the analytic and approximate 
template families at $10-120M_\odot$.  In \cite{ajith-2007}, a phenomenological
family of waveforms was proposed to model the coalescence of the \bbh{s} using
a hybrid method that combines analytical and numerical relativity waveforms.  
They achieve matches greater than 0.99 at $30M_\odot-130M_\odot$ for non-spinning \bbh{s}. 

In this paper, we conduct a first-step investigation into the 
ramifications on matched filtering of including spin in the \bbh{} 
waveforms provided by numerical relativity.
The set of spinning \bbh{} waveforms that we use are provided by the Penn State
group and were published in \cite{Herrmann07} in addition to several non-spinning
\bbh{} waveforms.  
The spinning binaries are axial configurations 
with the initial black holes having spins of equal magnitude where one is 
aligned and one anti-aligned with the orbital angular momentum for a set of
four spin parameters, $a=J/M^2=\{0.2,0.4,0.6,0.8\}$. 
The physical parameters of the waveforms are the
numerical resolution, $\Delta$, and the initial angular momentum of the black holes
in terms of the spin parameter, $a$.
Because the configuration is
axial,  the intrinsic parameters of the waveforms do not include
the angles between the spins and the axis of the orbital angular momentum.
We also include the radiation modes given 
by $\ell$ and $m$ of the spin weighted  $s=-2$, spherical harmonics, ${}_{-2}Y_{\ell m}$,
which are the inclination and azimuthal angles between the source and detector in the source
frame.  These waveforms contain approximately two and a half orbits.  Because of this, we 
restrict the total mass range to $m>50M_\odot$, set
by the initial frequency of the evolution and our choice of how we enter the LIGO band.

Both numerical accuracy, e.g.~truncation errors, and astrophysical
accuracy, e.g.~initial data choices, will play a role 
in determining the viability of numerical relativity \bbh{} waveforms
acting as templates.  Requirements  may be more stringent when characterizing the
sources of gravitational waves; however,  in this paper 
we focus on the use of numerical relativity waveforms as potential templates
for detection, not for parameter estimation.
Requirements for detection were first placed on waveforms from 
\bbh{s} generated by numerical relativity in references \cite{FHI, FHII}.  
This early work preceded the successful solution
of the \bbh{} problem by many years, but acts as a guide 
for determining the constraint on  numerical resolution accurate 
enough for data analysis purposes.  A similar method was 
employed more recently in reference \cite{Brady-2006}
in connection with \bbh{} evolutions of equal-mass,
non-spinning black holes over several orbits.
Ref. \cite{Brady-2006}, 
made a prediction of a maximum match that numerical waveforms will resolve.  
In addition, we did a preliminary
study of the impact that numerical errors can have on the faithfulness
of numerical waveforms using a Zerilli-based toy model in
\cite{BirjooLISA}.
We now employ similar tests as \cite{FHI} and \cite{Brady-2006} and verify
the predicted behavior of the match with resolution for our \bbh{} evolutions.

Often in numerical relativity, waveforms are extracted in
terms of coefficients of ${}_{-2}Y_{\ell m}$
the dominant mode being the quadrupole mode ($\ell=2$, $|m|=2$).
For compact binary inspiral searches, restricted post-Newtonian templates 
are commonly used for detection \cite{cutler-1993-70}. 
These templates include only the dominant 
harmonic in the amplitude while including as much information as possible
in the phase since phasing is the more important issue 
in matched filtering.  Corrected-amplitude templates have been 
considered in \cite{VanDenBroeck} and found to reduce the signal-to-noise
ratios for LIGO and add features to the detection and parameter
estimation for Advanced LIGO \cite{broeck-2007,broeck-2007-24}.

In \cite{Brady-2006}, it was found that using the $\ell=2$ mode was
was good enough for detecting gravitational waves of non-spinning, 
equal mass binaries.  
We include in our analysis the angle from the spherical harmonics, $\theta$ and $\phi$,
where $\theta$ is related to the inclination angle of the binary with respect to the 
detector but in the source frame.
We find that, in general, 
we will need to include higher modes to accurately represent spinning
\bbh{} waveforms. We also explore the sensitivity
of the match with the spin parameter of the black-holes' initial angular momentum
and how well a reduced template bank would do in matching with a
target template of arbitrary spin and inclination.

The paper is organized as follows.
In $\S$\ref{sec:techniques} we explain
the techniques we have employed in our \bbh{} code and in our data
analysis algorithms used in the rest of the paper.
We present our results in $\S$\ref{sec:results} including a description
of the equal-mass, spinning and non-spinning evolutions 
that produced the waveforms under consideration including their
spectra.  We first assess the quality of our waveforms in terms of resolution
through an analysis of the convergence properties of our \bbh{} code
in $\S$\ref{sec:convergence}.
We also  present how the match statistic can
reflect the convergence properties of our waveform in $\S$\ref{sec:resolution}.  
In $\S$\ref{sec:inclination} 
we present our study of the spinning templates in terms of 
varying inclination angle. 
Finally, in $\S$\ref{sec:comparisons} we look at using a specific numerical relativity
waveform as the template and check its ability to match a  target of unknown spin and inclination. 
We conclude in $\S$\ref{sec:conclusion}.

\section{Techniques} 
\label{sec:techniques}
\subsection{Numerical Relativity Techniques} 
We solve the fully nonlinear Einstein equations in the 3+1 form.
Bowen-York~\cite{BowenYork1980} puncture~\cite{punctures} initial data
representing the two black holes is specified on a spacelike slice $t
= 0$.  This allows us to specify the initial mass, linear momentum,
coordinate location and spin of each black hole.  The
BSSN~\cite{Nakamura87, Shibata95, baumgarte_shapiro} formulation of
the Einstein equations in the $\chi$ form~\cite{Campanelli05a} is used
to evolve this data forward in time.  The gauge conditions are
modified versions of the 1+log lapse and $\Gamma$-driver shifts,
\begin{eqnarray}
\partial_0 \alpha &=& -2\alpha K \, ,\\
\partial_0 \beta^i &=& 3/4 B^i \, , \\
\partial_0 B^i &=& \partial_0 \tilde{\Gamma}^i- \eta B^i \, ,
\end{eqnarray}
where $\partial_0 = \partial_t - \beta^j \partial_j$.  We choose $\eta
= 2$ as in \cite{JenaCalibration2006}.  The advection term
$\beta^j\partial_j\tilde{\Gamma}^i$ removes certain zero-speed modes
of the system as analyzed in~\cite{vanmeter-2006-73}. These gauge
choices, with the exception of this advection term, were found to be
important for long-term stable and accurate evolutions of head-on
collisions without excision~\cite{Alcubierre02a}.  The puncture
locations are determined by integrating the equation $\partial_t
x^i(t) = \beta^i(x(t))$ \cite{CampanelliPRL2006}.

The Weyl scalar $\Psi_4 = R_{\alpha \beta \gamma \delta} n^\alpha \bar
m^\beta n^\gamma \bar m^\delta$ measures outgoing gravitational
radiation at large distances from the source.  The orthonormal tetrad
$\{n^\alpha, l^\alpha, m^\alpha, \bar m^\alpha \}$ is constructed from
the spherical coordinate basis vectors as in \cite{lazarus}. We
decompose $\Psi_4$ into ${}_{-2}Y_{\ell m}$ on
spheres of constant coordinate radius.

We use the Cactus infrastructure for parallelization, I/O and
parameter handling.  The initial data is computed using the {\em
TwoPunctures}~\cite{ansorg} thorn, a spectral code which solves the
momentum constraint for the conformal factor of our conformally flat
spatial metric.  We use the {\em moving punctures} approach without
excision~\cite{Baker05a,Campanelli05a} so the singularity is not
treated specially, apart from ensuring that it does not lie on a grid
point initially.  The evolution is performed using our BSSN thorn
which is automatically generated using the \texttt{Kranc}~\cite{Kranc}
code generation package.  $\Psi_4$ is also computed using a
Kranc-generated thorn.  For mesh refinement we use
\texttt{Carpet}~\cite{erikfmr}, and our \texttt{MoveIt}
infrastructure for specifying the grid structure based on the
locations of the black holes.

The code uses finite differences, and all spatial derivatives are
computed using centered fourth order stencils, apart from the shift
advection terms which are computed using upwind, lop-sided, fourth
order stencils.  This gives increased accuracy over using centered
stencils throughout.  The evolution stencil thus has three points
adjacent to any given point in all three directions. For time
integration we use fourth order Runge-Kutta with a Courant factor
$dt/dx^i = 1/2$.

We use 9 levels of box-in-box mesh refinement, where the outermost
(base) grid covers the domain $x^i \in [-320, 320]$.  This allows the
outer boundary to be causally disconnected from the computation of
$\Psi_4$ at $r = 30m$ until well after the wave has passed.  The next
three grids cover the domains $x^i \in [-160, 160]$, $[-80, 80]$, and
$[-40, 40]$. These grids are all fixed in place throughout the
simulation.  The remaining 5 levels generically contain two grids each
centered on one of the black holes.  These grids are cubical of
half-side $10$, $5$, $2.5$, $1.25$ and $0.625$.  Our moving grids
infrastructure handles the case of the two grids on a given refinement
level overlapping by replacing them with a single grid which is the
smallest grid containing both of them.  Each grid has half the grid
spacing of its parent grid, and we typically refer to the resolution
of the finest grid as $h_f$.  This grid structure allows the
extraction sphere at $r = 30m$ to not intersect any refinement
boundaries.

We use Berger-Oliger mesh refinement as implemented by
\texttt{Carpet}, and we refine in time as well as in space.  The
evolution on the refined grids requires a boundary condition provided
by the coarse grid.  This is implemented by adding additional points
outside the refinement boundary, which are filled by interpolation
from the coarser grid.  Since the evolution stencil is of size 3, this
boundary must consist of at least three points.  Since the fine grid
must be updated twice as frequently as the coarser grid, the
interpolation is in time as well as space.  As noted in
\cite{erikfmr}, for evolution equations which are first
order in time but second order in space, this scheme leads to
instabilities.  One solution is to interpolate from the coarse grid
only at the start of the Runge-Kutta sub-stepping algorithm, not at
every sub-step.  Since on each sub-step the boundary points are not
computed, additional {\em buffer points} are required outside the
finer grid.  These points are also filled by interpolation at the
start of the sub-stepping algorithm.  The total number of boundary
points required is $\mbox{stencil\_size} \times
\mbox{time\_integrator\_sub-steps} = 12$.  We use fifth order
interpolation in space, and second order interpolation in time.  This
lowers the overall accuracy expected from 4th to 2nd, but using higher
order time interpolation is prohibitively expensive in terms of
computational resources.  We find that using 9 boundary points instead
of 12 does not affect the results significantly, but it reduces the
computational cost \cite{JenaCalibration2006}.  We expect that the second order error from the
mesh refinement boundaries is small, and that at low resolutions the
scheme will appear fourth order accurate.

The initial data parameters are given in Tab.~\ref{tab:initialdata}.
The non-spinning R1 parameters are taken from \cite{NASA2006BBH},
based on the quasi-circular sequence found in \cite{CookID,
BaumgarteID, LazarusID}.  The spinning runs are those reported in
\cite{Herrmann07}.  These runs
consist of two equal-mass black holes whose initial spins are oriented
such that one is aligned with the orbital angular momentum and the
other is anti-aligned, and the magnitudes of the spins are equal.  The
runs are labeled as S0.05-S0.20 and have spin parameters
$a=\{0.2,0.4,0.6,0.8\}$.  
The black holes are located at positions $(0,\pm
y,0)$, have linear momentum $(\mp P,0,0)$, spin $(0,0,\pm S)$, and
bare puncture masses $m_{\pm}$.  The irreduceable masses (measured
from the areas of the apparent horizons) are $m_1 = m_2 = m/2$ where
$m=m_1+m_2$ is the total mass of the binary.

\begin{table}[h]
\caption{Initial data parameters}
\begin{center}
\begin{tabular}{l c c c c c c c}
\hline   
Model & $y/m$ & $P/m$ & $a$
& $m_+/m$ & $m_-/m$ & $m_\textrm{ADM}/m$ & $J_\textrm{ADM}/m^2$ \\
 \hline 
 \hline
R1     & 3.257 & 0.133   & 0    &  0.483  & 0.483  & 0.996   & 0.868 \\
S0.05  & 2.95  & 0.13983 & 0.2  &  0.4683 & 0.4685 & 0.98445 & 0.825 \\
S0.10  & 2.98  & 0.13842 & 0.4  &  0.4436 & 0.4438 & 0.98455 & 0.825 \\
S0.15  & 3.05  & 0.13547 & 0.6  &  0.3951 & 0.3953 & 0.98473 & 0.825 \\
S0.20  & 3.15  & 0.13095 & 0.8  &  0.2968 & 0.2970 & 0.98499 & 0.825 \\
 \hline
\end{tabular}
\end{center}
\label{tab:initialdata}
\end{table}        

\subsection{Data Analysis Techniques}  
Our numerical waveforms are calculated during evolution in terms of
the Newman-Penrose $\Psi_4$, where
\begin{equation}
r m \Psi_4(t,\vec r) = \sum_{lm} {}_{-2} C_{lm}(t,r)
{}_{-2}Y_{lm}(\theta,\phi) \,,
\end{equation}
and $r$ is the extraction radius.
The relationship between $\Psi_4$ and $h_+(t)$ and $h_\times (t)$ is
given by
\begin{equation}
\Psi_4(t) = \frac{d^2}{dt^2}(h_+(t) - i h_\times(t))\, .
\end{equation} 
To carry out the matched filtering analysis, we compute $\tilde h_+ (f)$ and $\tilde h_\times (f)$
directly from the Fourier transform of the real part of $\Psi_4$ such
that
\begin{equation}
\tilde h_+ (f) = \mathcal F(\mathrm{Re}(\Psi_4))(f)/(-4 \pi^2 f^2) \,.
\end{equation} 
This avoids issues regarding integration constants that arise from the
time domain integration of $\Psi_4$ \cite{BertiJENA}.  
Our initial data contains gravitational radiation which does not correspond to
that present in an astrophysical situation.  This radiates away and is
visible in the waveform as an {\em initial data pulse}.  We remove
this pulse from each waveform.

Our analysis of the match of the numerical relativity waveforms will
follow along the lines of the matched filtering procedure for
detecting gravitational waves.  That is, given two time domain
waveforms $h_{1}(t)$ and $h_{2}(t)$, the scalar or inner product
between these two functions is defined as
\begin{equation}
\label{eq:inner}
\langle h_1 | h_2 \rangle = 4 \, \mbox{Re}\int_\fmin^\fmax \frac{\tilde
h_1(f) \tilde h^*_2(f)}{S_{h}(f)} \, df,
\end{equation}
where the domain $[\fmin,\fmax]$ is determined by the detector 
bandwidth and $\tilde h(f)$ stands for the Fourier transform of
the respective time series.  $S_h(f)$ denotes the noise spectrum for
which we use the initial LIGO noise curve.   The fact that our numerical
waveforms only contain a few orbits before merger means that they do 
not span the entire LIGO frequency band.  The initial 
orbital frequency of these runs
varies depending on the value of the spin
between approximately $0.016/m$ and $0.024/m$.
The most stringent lower limit on $\fmin$ would
be $0.024/m$.  We impose the condition that the signal-to-noise ratio would
be coming entirely from the domain spanned by our numerical waveforms
such that $\fmin$ is calculated for each mass of the template.
This fixes the lowest mass for which the templates could
be useful to $50M_\odot$.

The match statistic \cite{Owen-96}, is properly defined as the maximized overlap between the
signal and the template, however, we will use it as a measure between
two templates $h_1$ and $h_2$.  There are two  extrinsic 
parameters: the time of arrival of the signal, $t_0$, and the initial phase of the orbit
when it enters the LIGO band, $\Phi$. First, consider
the maximized overlap between templates defined with 
only maximization over time given by
\begin{eqnarray}
 O_{max}[h_1,h_2] &\equiv& \max_{t_0} O[h_1,h_2]\, ,\label{eqn:match}\\
 O[h_1,h_2] &\equiv& \frac{ \langle h_1 |h_2
\rangle} {\sqrt{\langle h_1 | h_1 \rangle \langle h_2 | h_2 \rangle
}}\,.
\end{eqnarray}  
In Eq.~(\ref{eqn:match}), the maximization is to be understood as the
maximum overlap between $h_1$ and $h_2$ obtained by shifting the
template in time, i.e. $h(t) \rightarrow h(t+t_0)$ leading to the
numerator being written in frequency domain as
\begin{equation}
\max_{t_0} \, \langle h_1 | h_2 \rangle = \max_{t_0} 4 \mathrm{Re} \,
 \int_{\fmin}^{\fmax} \frac{ \tilde h_1(f) \tilde h_2^*(f) e^{2
 \pi i f t_0}} {S_h (f)} \, df\,.
\end{equation}
The Fourier transform is replaced by a discrete fast Fourier transform, transforming the
integral into a discrete sum.  When computing the match without phase
optimization, we will typically compute the match of the templates
using only $h_+$.  

We also compute the match
with an optimization over the phase, $\Phi$ of the template
in addition to $t_0$. A
waveform of arbitrary initial phase $\Phi$ is written as
\begin{equation}
\tilde h(f)=\tilde h_+ (f) \cos \Phi + \tilde h_\times (f) \sin \Phi \,.
\end{equation}
Given an arbitrary waveform, $\tilde h_1$, the phase optimization over
the template, $\tilde h_2$, can be carried out using the normalized templates
$e_{i,+,\times} = \tilde h_{i +,\times }/||\tilde h_{i +,\times}||$ where $i$ runs
over ${1,2}$. In reference
to the match we will always be referring to frequency domain templates and 
for ease of notation we shall refer to $\tilde h$ as $h$ in the match formulae.
The {\em typical} match \cite{mohanty-1998-57} is given by
\begin{eqnarray}
\Mtyp &\equiv& \max_{t_0} \, \max_{\Phi_2} O[h_{1+},h_2] \nonumber \\
      &\approx& \max_{t_0} \sqrt{O[e_{1+},e_{2+}]^2
+ O[e_{1+},e_{2 {\times }}]^2}\,,
\label{eqn:oldtypmatch}
\end{eqnarray}
where we have assumed that $e_{2+}$ and $e_{2\times }$ are nearly orthogonal,
i.e.~$\langle e_{2+}|e_{2\times} \rangle \approx 0$.  This
approximation is valid only for angles near ($\theta = 0$, $\phi=0$)
where the contribution from the modes other than $\ell=|m|=2$ is zero
because of the spherical harmonic values. For a general waveform at
nonzero $\theta$ and $\phi$, the overlap of the
corresponding $h_+$ and $h_\times$ is of the order of a few percent.
In our spinning waveforms, we find a maximum deviation from orthogonality
of $\sim 3\%$ at an angle that approaches $\pi/2$.
In order to avoid uncertainties
in the matches at large angles, we construct an orthonormal waveform basis as
done by \cite{sathyaprakash-1991, mohanty-1998-57} and outlined below.

Any arbitrary
polarization other than $(+,\times)$ can be expressed as a linear
combination of these linearly independent basis vectors. The phase
optimization of the match can then be done in terms
of the new orthonormal basis vectors,  one of which we choose to be $e_{i +}$
and the other $e_{i \perp}$  given as
\begin{equation}
e_{i \perp}=\left( e_{i \times} - e_{i +}  \langle e_{i +} |
e_{i \times} \rangle \right) (1- \langle e_{i +}| e_{i \times}
\rangle^2)^{-1/2}\,.
\end{equation}
One can see that $\langle e_{i +}| e_{i \perp} \rangle = 0 $ by
construction.  Given such orthonormalized basis vectors for two sets
of parameters $i = 1, 2 $, 
we can calculate the upper and  lower bounds on the phase optimized match
\cite{damour-1998-57}.

In this new orthonormal basis,  we rewrite the typical match from Eq.~(\ref{eqn:oldtypmatch})
by matching the $+$ polarization of one template while optimizing over the phase of the other. 
This mimics the situation in which one of the waveforms acts as the template and can 
be maximized over its phase, $\Phi_2$, while keeping the phase of the second template,
$\Phi_1$, fixed, as follows
\begin{eqnarray}
\Mtyp  &\equiv&  \max_{t_0} \, \max_{\Phi_2} O[h_{1+} ,h_2 ] \nonumber \\
&=& \max_{t_0} \sqrt{O[e_{1+},e_{2+}]^2
+ O[e_{1+},e_{2 \perp}]^2}\,. 
\label{eqn:typical}
\end{eqnarray}
In general, the phase of the signal could take any value, and one would like to 
know the best and worst possible values of the match.  
The expression for the {\em best} 
match (the upper bound) is given by \cite{damour-1998-57}
\begin{eqnarray}
\Mbest &\equiv& \max_{t_0}\, \max_{\Phi_1} \, \max_{\Phi_2} O[h_1,h_2 ]  \\
&=& \max_{t_0} \,\left[ \frac{A+B}{2} +
\left[ \left( \frac{A-B}{2}\right)^2 + C^2 \right]^{\frac{1}{2}} \nonumber
\right]^{\frac{1}{2}}
\label{eqn:best}
\end{eqnarray}
in which the phases of each template are optimized. 
The {\em minimax}  match is given by the case when 
one maximizes the phase of one of the templates but minimizes over the other. This is to 
mimic the worst case scenario when the signal phase is such that it gives lower matches
even when maximized over the template phase, given by \cite{damour-1998-57}
\begin{eqnarray}
\Mminimax &\equiv&  \max_{t_0}  \min_{\Phi_2} \max_{\Phi_1} O[h_1 ,h_2 ] \\
&=&  \max_{t_0}\,  \left[
\frac{A+B}{2} - \left[ \left( \frac{A-B}{2}\right)^2 + C^2
\right]^{\frac{1}{2}} \right]^{\frac{1}{2}} \nonumber
\label{eqn:minimax}
\end{eqnarray}
and for both cases the functionals $A,\ B, \ C$ are written in terms of the orthonormal
basis functions $(e_{1+}, e_{1\perp})$ and $(e_{2+},
e_{2 \perp})$ corresponding to the two templates being compared
\begin{eqnarray}
A & \equiv & \langle e_{1+} | e_{2+} \rangle^2 + \langle
e_{1+} | e_{2 \perp} \rangle^2 \, ,\nonumber \\ 
B & \equiv & \langle
e_{1 \perp} | e_{2+} \rangle^2 + \langle e_{1 \perp} | e_{2 \perp}
\rangle^2 \, ,\nonumber \\ 
C & \equiv & \langle e_{1+} | e_{2+}
\rangle \, \langle e_{1 \perp} | e_{2+} \rangle + \langle
e_{1+}| e_{2 \perp} \rangle \, \langle e_{1 \perp}| e_{2 \perp} \rangle \,. \nonumber 
\end{eqnarray}

Note the symmetry of the formulae upon interchange of  the two templates
which arises from the fact that only the {\it relative phase} between the two
templates should matter.
Because of this, which template is maximized and which minimized is interchangeable.
We will refer to the target's phase being minimized 
and the template's phase  being maximized since this makes sense in a detection scenario.

\section{Results}
\label{sec:results}
\begin{figure}[h]
\includegraphics[height=8cm]{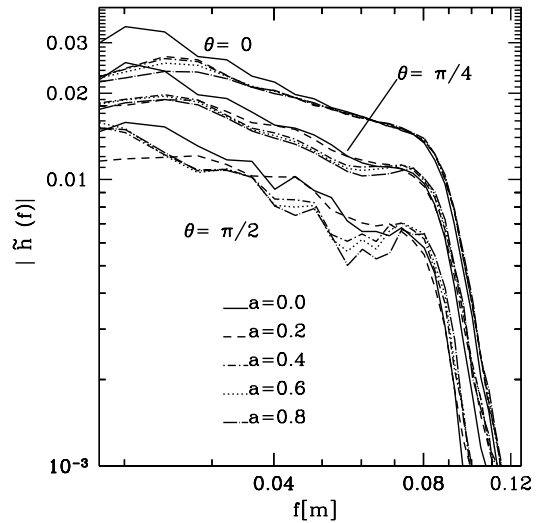}
\caption{We plot the spectrum of the wave versus frequency
of waveforms from different initial spins
 in a log-log plot.  The spinning cases all have
resolutions of $m/40$ and the non-spinning, $m/38.4$.  
We present three cases of inclination angle, $\theta$, for each set of spins, and 
the units are $f_{\mathrm{code}}=f_{\mathrm{phys}}/m$.}
\label{fig:fftspectra}
\end{figure}

We present our results in three categories: the variation of the match with resolution,
including convergence tests of the waveforms; the variation of the match with inclination 
and spin; and how well we can differentiate between a target of unknown inclination
angle and spin in the mass range of $50M_\odot$ to $300M_\odot$ given a template of fixed parameters.
The spectra of the \bbh{} waveforms from the models R1 and S0.05-S0.20
are shown in Fig.~\ref{fig:fftspectra}, in which we plot $|h(f)|$.
The $\theta=0$ lines correspond to the $\ell=|m|=2$ mode.
The final black hole in all these runs settled to the same
final spin with a spin parameter $a\approx 0.66$. 

In Fig.~\ref{fig:fftspectra}, we can see that when $\theta=0$, the spectra 
for each value of $a$ become very similar for $f>0.05m$.  This is because
all of the configurations we evolved settled down to very similar final black holes.
When $\theta$ deviates from zero, however, we can already see from the spectra that 
we get more variation between the different spinning waveforms.  This is in part
due to the larger resolution requirements for the higher harmonics, however, as we shall
see in the following results, the resolution differences do not account for all of the variation.

Firstly, in $\S$\ref{sec:convergence}, we investigate the convergence of the dominant,
$\ell=|m|=2$, mode for non-spinning
\bbh{s}, the convergence of the spinning waveforms was published in 
\cite{Herrmann07} to be between third and fourth order.
In $\S$\ref{sec:resolution} we investigate the quality
of the numerical waveforms as templates in matched filtering
in terms of the resolution including spin and inclination angle.
Next, in $\S$\ref{sec:inclination}, we show how the inclusion of modes
$\ell <5$ affects the match calculations when compared to just using the
dominant mode waveform $\ell=|m|=2$ as a function of the spin.  
Additionally, we determine the effect of using a finite 
extraction radius on the quality of the matches.
Lastly, in $\S$\ref{sec:comparisons}, we compare 
the different spin waveforms to each other, including
the dependence on $\theta$ but keep the masses the same.
This is equivalent to checking the {\em faithfulness}
\cite{damour-1998-57} of
the waveforms assuming the orientation of the source for the different
sources to be the same. While this assumption will need to be relaxed
in a fully general treatment for data analysis, it suffices to
demonstrate the importance and impact of using higher than dominant
modes in the analysis of these merger waveforms.

\subsection{Numerical Convergence}
\label{sec:convergence}
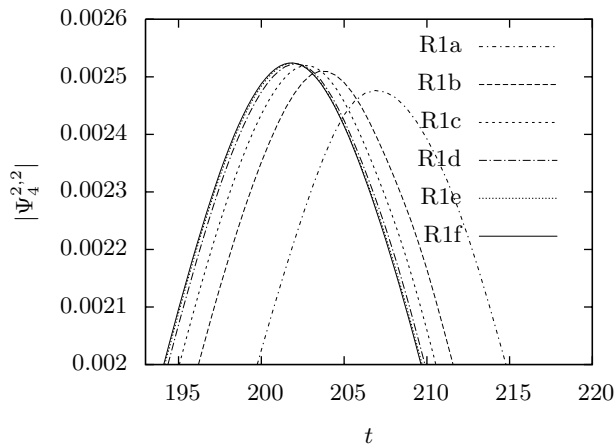
\begin{figure}[h]
\input{amp-all-include.tex}
\caption{Monotonic change in $|\Psi_4^{2,2}|$ as resolution is
increased.  It appears that the function is converging to a continuum
solution.}
\label{fig:waveamps}
\end{figure}

A convergence test of the spinning series of runs called S0.05-S0.20 is 
given in \cite{Herrmann07}.
Here we investigate of the quality of our numerical waveforms
with a convergence test of the equal-mass, non-spinning R1 series of runs.
These runs are names R1a-R1f and only differ in their resolution.  
The suffix corresponds to the grid spacing of the finest grid surrounding
each black hole.  The grid spacings  are
 $\Delta_{a-f} =\{m/25, m/32, m/38.4, m/44.8, m/51.2, m/57.6\}$. 
We consider the
convergence properties of the $\ell =m = 2$ mode of $\Psi_4$ computed
on the coordinate sphere at $r = 30$.  In Fig.~\ref{fig:wavereal} we
plot $\mathrm{Re}[\Psi_4^{2,2}]$, and Fig.~\ref{fig:waveamps} shows
that $\Psi_4^{2,2}$ computed at $r = 30$ appears to converge
monotonically with resolution to a continuum solution, and that the
high resolution results agree well with each other.  Similar results
(not shown) are obtained for $\arg \Psi_4^{2,2}$ and the coordinate
locations of the black holes.  Using the runs R1a, R1b, and R1f we
demonstrate fourth order convergence of the amplitude and phase of
$\Psi_4^{2,2}$ in Fig.~\ref{fig:waveconvamp} and
Fig.~\ref{fig:waveconvphase}.  Were we to plot only the highest resolution
runs, the fourth order convergence would be lost; i.e.~there is a
source of error which spoils the convergence at high resolutions.
However, as can be seen from Fig.~\ref{fig:waveamps}, the waveforms
still appear to converge to a continuum solution, just at a different
order.

\begin{figure}[h]
\input{convergence-re-include.tex}
\caption{$\mathrm{Re}[\Psi_4^{2,2}]$ computed at $r = 30$ for three
different resolutions.  The feature at $t = 40$ is due to the
gravitational radiation present in the initial data, and will be cut
out of the waveform before taking Fourier transforms for data
analysis. }
\label{fig:wavereal}
\end{figure}
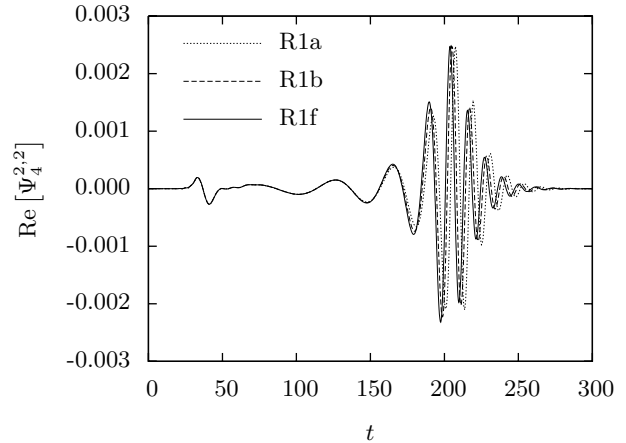

\begin{figure}[h]
\input{convergence-amp-include.tex}
\caption{Convergence of $|\Psi_4^{2,2}|$ using the highest and the two
lowest resolutions.  $C_3, C_4, C_5$ are the scaling factors expected
for third, fourth and fifth order convergence respectively.  We see
that $a-b$ matches most closely with $C_4 \times (b - f)$ indicating
fourth order convergence. }
\label{fig:waveconvamp}
\end{figure}
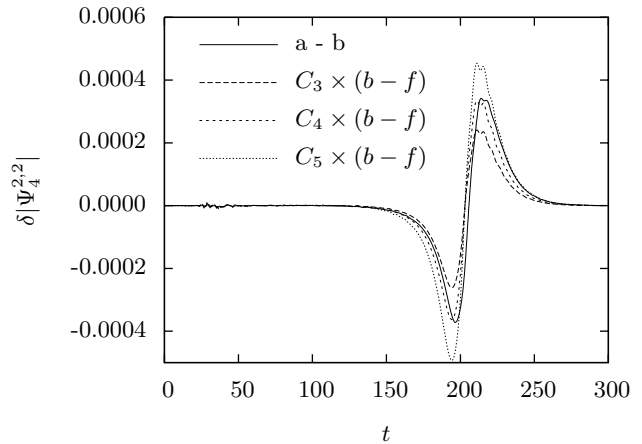

\begin{figure}[h]
\input{convergence-phase-include.tex}
\caption{Convergence of $\arg \Psi_4^{2,2}$.  As in
Fig.~\ref{fig:waveconvamp}, we see fourth order convergence.}
\label{fig:waveconvphase}
\end{figure}
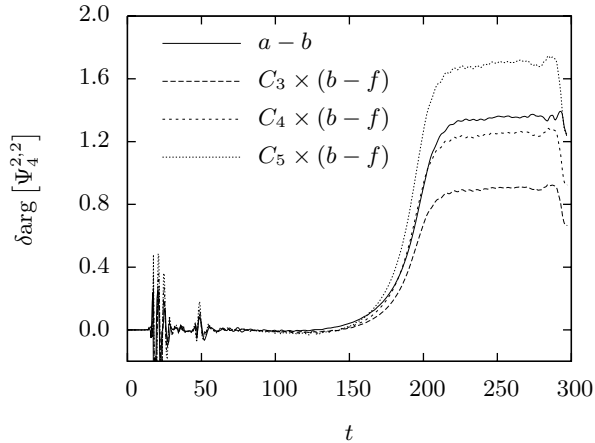

\subsection{Variation of Templates with Resolution}
\label{sec:resolution}
We can follow the numerical convergence properties through 
matched filtering since we know how the numerical
codes approximate the exact solution to the partial differential 
equations discretely. For finite differencing this is expressed as
\begin{equation}
h(t) = h_0(t) + c(t)\Delta^p + \mathcal O(\Delta^{p+1})
\label{eqn:discrete}
\end{equation}
where $h_0(t)$ is the exact solution found when the resolution goes to infinity, $\Delta$ is the
grid spacing ($\Delta=\Delta x=\Delta y=\Delta z\propto \Delta t$),
$p$ is the order of the truncation error and $c(t)$ is a time dependent
scaling independent of the grid spacing.  Since our code is
approximately fourth order accurate, we expect to scale with $p=4$.
One can calculate $h_0(t)$ using Richardson extrapolation of three discrete, convergent
solutions of different resolution.
We have computed $h_0(t)$ for the R1 series of runs using R1a, R1b, and R1f and
found that $h_0(t)$ 
coincides to within $10^{-5}$ in amplitude with the waveform evolved with the 
finest resolution, $\Delta_f=m/57.6$. 
In light of this, we will use our finest resolution run, indicated as $h(\Delta_f)$, 
in place of the Fourier transform of $h_0(t)$.
  
\begin{figure}[h]
\includegraphics[height=7cm]{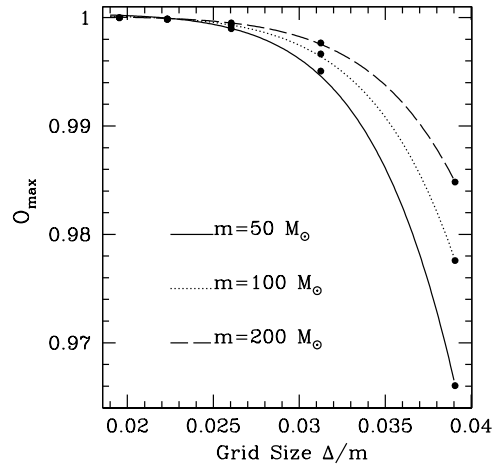}
\caption{The maximized overlap, $\Omax[h_i,h_f]$, from Eq.~(\ref{eqn:match}) 
is plotted as a function of grid size for three total masses indicating fourth
order convergence. The data are given by the points and the least square fitting by the lines.
The $p=4$ fit differs for different masses because of the weighting by the 
current LIGO noise curve $S_{h}(f)$. }
\label{fig:Mvsgrid4}
\end{figure}

We can predict the behavior of the maximized overlap as a function of resolution 
between two waveforms that differ only in their
resolution  by expanding the match equation, Eq.(\ref{eqn:match}), about $\Delta^p=0$.  
Note that when $\Delta^p$ is a constant in time, we can
express Eq.(\ref{eqn:discrete}), in the Fourier space as $h_i = h_0
+ c \Delta^p_i$, where $i$ is running over each waveform.  
We expand the match for the case of two templates at two different
resolutions to be as general as possible, i.e. $O[h_1(\Delta_i),h_2(\Delta_j)]$ 
as follows
\begin{eqnarray}
O[h_{i},h_{j}] &=& 1- \frac{1}{2}(\Delta^p_{i}-\Delta^p_{j})^2
 \left( \frac{ \langle c|c \rangle} { \langle 
 h_0| h_0 \rangle } - \left(\frac{ \langle  c| h_0
 \rangle^2} { \langle h_0| h_0\rangle ^{2}} \right )
 \right) \nonumber  \\
 & + & \mathcal O^{(3)}(\Delta^p)\,.
\label{eqn:expandedmatch}
\end{eqnarray}
The above equation was expanded for two matches of different resolution
but the match with the Richardson extrapolated solution can be recovered 
by setting $\Delta_j=0$ up to the order expressed.
This equation indicates that the mismatch, ($1-\Omax$), goes like $\Delta^{2p}$.
Fig.~(\ref{fig:Mvsgrid4}) shows the dependence of the $\Omax$ on the
resolution for three choices of the total mass, $m = \{50,100,200\}M_\odot$.  
The least squares fit is done for several possible
values of $\Delta^{2p}$ including $p=3,4,5$. The best fit is
found for $p=4$,  and is the fit pictured in Fig.(\ref{fig:Mvsgrid4}).

Because of the computational cost of generating
solutions to the Einstein equations at three resolutions for 
every waveform of interest, Flanagan and Hughes \cite{FHII} suggested
computing the match between templates of neighboring resolutions,
such as $\Omax[h(\Delta_a),h(\Delta_b)]$ for example.
A variation, used by Baumgarte et al \cite{Brady-2006}, is to compute
the series of matches as $\Omax[h(\Delta_i),h(\Delta_f)]$, where the first
template runs over all the resolutions available, denoted by $\Delta_i$,
and the second template's resolution is fixed to the finest available.

\begin{figure}[h]
\includegraphics[height=8cm]{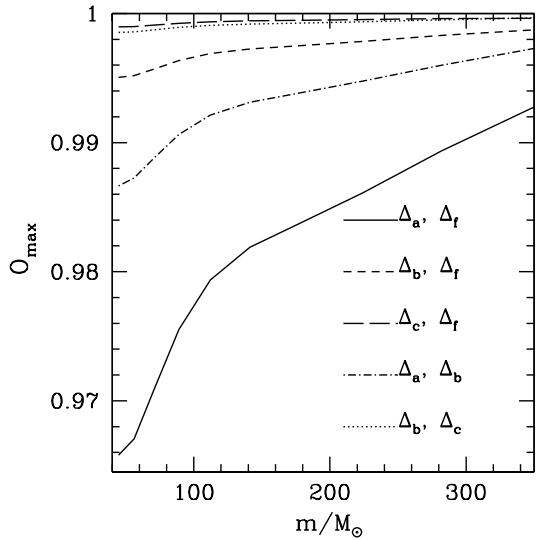}
\caption{The maximized overlaps, $\Omax[h_{1+},h_{2+}]$,
are plotted as a function of mass for several resolutions of our equal-mass \bbh{} series of
runs.  Labels $(a,b,c)$ stand for the coarse runs
$R1a,b,c$ with resolutions $m/25.6$, $m/32.0$, $m/38.4$, while $R1f$ has resolution 
$m/57.6$. 
The overlap between adjacent resolutions $\Omax[h_{+}(\Delta_a),h_{+}(\Delta_b)$ 
has deceptively high values compared to the matches with the finest resolution.}
\label{fig:Mvsmass}
\end{figure}

We plot the Flanagan and Hughes type of  overlaps,
$\Omax[h_+(\Delta_a),h_+(\Delta_b)]$ and 
$\Omax[h_+(\Delta_b),h_+(\Delta_c)]$ 
and three cases where we vary the resolution of one template while keeping the second
template fixed to the finest resolution, $\Omax[h_+(\Delta_i),h_+(\Delta_f)]$ where
$i=a,b,c$, in Fig.~(\ref{fig:Mvsmass}). These matches are unoptimized over the  phase,
but optimized over the time $t_0$ and follow the definition given in Eq.~(\ref{eqn:match}).

The threshold of match above which the resolutions will be
"good enough" for detection is typically set to 0.98.  
Clearly from the figure we can conclude for the $\theta=0$, a resolution
of $\Delta_b=m/32.0$ was enough for the entire mass range to reach a minimum match of
0.98 and that, for a lower resolution of $\Delta_a=m/25.6$, 
we only achieved $\Omax > 0.98$
for total masses larger than $100M_\odot$.
One of the main features of the matches with resolution is the decrease in the match toward 
smaller masses.  This is partly due to the low number of gravitational wave cycles
we have present in the waveforms. The numerical errors, however, will increase as we 
increase the total number of cycles so the choice of resolution will depend on the number
of orbits evolved.  As we evolve more cycles, we will also be extending the total
mass range into smaller masses.

We now have to choose which match to use in assessing the quality of the templates.
The matches are expected to get closer to one when phase optimization is included.
Since the phase of a given signal waveform depends on the detector orientation and several such 
variables, it could differ from the template up to a constant phase factor and the match would vary 
depending on what the phase is. The worst case would occur when the arbitrary 
phase is such that there is maximum destructive interference between the
target and the template, and sets up the lower bound on the possible matches, $\Mminimax$. If one does
not add any phase factor to the numerically generated template , one gets what is called the
typical match and if the phase factor is chosen for maximum constructive interference,
it yields the "best" match, $\Mbest$.  The value of $\Mbest$ is a measure of the 
closest distance between two templates. We'll use the minimax match for the rest of the paper to 
report our results in terms of lower bounds and use the typical and best matches
to illustrate the range over the different type of  matches. 

To demonstrate the differences in the match with choice of optimization, we tabulate examples
of optimization in Tab.~\ref{tab:optim} for few selected total masses 
between the coarsest and finest runs.
No optimization at all is listed in the second column, 
optimization over just $t_0$ from Eq.~(\ref{eqn:match}) in the third, the typical match
Eq.~(\ref{eqn:typical}) in the fourth, minimax, Eq.~(\ref{eqn:minimax}),
in the fifth and finally best match, Eq.~(\ref{eqn:best}), in the final column.
The phase optimized matches, the last three columns, exceed
the the threshold of $\Omax > 0.98$ 
for even the least resolved waveform when matched with the 
finest resolution waveform.  
\begin{table}[h]
\caption{Comparing matches with different optimizations between the coarsest ($R1 a$) 
and the finest ($R1f$) runs for selected total masses. The second column
is the normalized overlap between the two waveforms without optimization over $t_0$,
the third column is the match with optimization over $t_0$, the fourth is the 
typical match, the fifth the minimax, and the final column is the best match.} 
\begin{center}
\begin{tabular}{c c c c c c}
\hline $m$ &  $O$& $\Omax$ & $\Mtyp$ & $\Mminimax$ & $\Mbest$\\
 \hline 
 \hline
$44.67$ & $0.7987$   & $0.9658$ & $0.9979$ & $0.9974$ & $0.9983$\\[1mm]
$70.79$ & $0.5458$   & $0.9709$ & $0.9989$ & $0.9989$ & $0.9991$\\[1mm]
$112.2$ & $0.2308$   & $0.9794$ & $0.9994$ & $0.9993$ & $0.9995$\\[1mm]
$177.8$ & $0.04864$  & $0.9837$ & $0.9994$ & $0.9993$ & $0.9996$\\[1mm]
$281.8$ & $-0.06719$ & $0.9893$ & $0.9993$ & $0.9993$ & $0.9996$\\[1mm]
   \hline
\end{tabular}
\end{center}
\label{tab:optim}
\end{table}        

We also calculated the minimax matches, see Eq.~(\ref{eqn:minimax}), for all the $R1$ resolutions,
although we do not plot them here, and found that for all the resolutions
$\Mminimax \ge 0.99$ in the mass range under consideration and show the same
variation as the overlap.  
These results hold only for $\theta=0$ and $a=0$. 

We now analyze the resolution needs of the spinning \bbh{} waveforms
of which we only have three resolutions for each of the spinning configurations
in contrast to the five available for the non-spinning case.  These are labeled
$\Delta_{\mathrm{coarse}}=m/32$, $\Delta_{\mathrm{med}}=m/38.4$, and $\Delta_{\mathrm{fine}}=m/40$. 
Later in this paper, we find that it is necessary to include the higher modes 
for the spinning \bbh{} templates; and, therefore,
as we analyze the spinning \bbh{} waveforms versus resolution we also include
its variation with $\theta$.
At inclination angle $\theta=0$ only the $\ell=|m|=2$ mode
is present in $h$, and when $\theta \neq 0$ all the modes get mixed.  
We do not include modes $\ell \ge 5$ in constructing $h$ because these were too small
to be well resolved for the evolutions we use in this paper. 
The variation of $\theta$ is presented for two suggestive cases, $\theta=\pi/4$ and $\theta\sim\pi/2$.
Values of $\theta$ are not taken to be exactly on the plane since the radiation is
then linearly polarized and cannot be described with two basis vectors and
cannot be maximized over the phase at that point, but instead at 
$\theta\sim \pi/2 = 89 \pi/180$.

\begin{figure}[h]
\includegraphics[height=7cm]{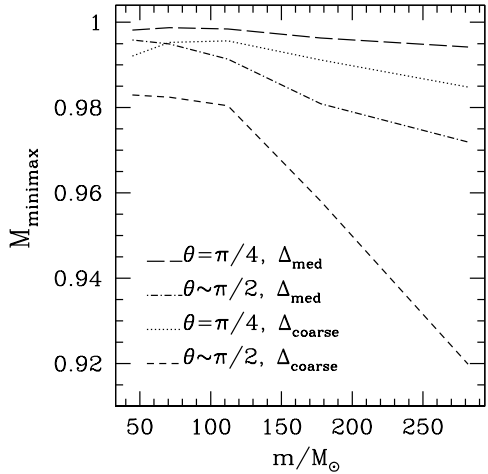}
\caption{We plot the minimax match between the
waveforms of $a=0.8$ for two cases of inclination angle, $\theta=\pi/4$ and $\theta=\pi/2$
and two different resolutions $\Delta_{\mathrm{fine}}$ with $\Delta_{\mathrm{med}}$ 
and $\Delta_{\mathrm{fine}}$ with $\Delta_{\mathrm{coarse}}$. 
Corresponding matches for lower spins are all $\ge 0.99$ and are not shown here.}
\label{fig:spin0.8reslns}
\end{figure}

For the $\theta=0$ case, all the spinning waveforms have a $\Mminimax \ge 0.99$
over the entire mass range for matches between the $\Delta_{\mathrm{coarse}}$ and 
$\Delta_{\mathrm{fine}}$ and between $\Delta_{\mathrm{med}}$ and $\Delta_{\mathrm{fine}}$.
These matches also follow the same trend with mass, i.e. 
decreasing matches with decreasing total mass,
as reported in Tab.~\ref{tab:optim}.
For $\theta \neq 0$, we plot the minimax match for the $a=0.8$ spin case
between both the highest and coarsest and highest and medium resolved waveforms
at the two values of $\theta$ in Fig.~\ref{fig:spin0.8reslns}.
We only show the $a=0.8$ waveforms since the matches of the high and low resolution of
the other spin and non-spinning runs are all $\Mminimax \ge 0.99$ at these angles again 
over the entire mass range.  Note, however, that even though the matches are very
high, they decrease with increasing spin.  This is expected since we are keeping the number
of points across the black hole fixed, but as the spin increases, the horizon area decreases,
hence the effective numerical resolution of a high initial spin waveform 
is lower than that of  a low initial spin
waveform.
The case of $\theta \sim \pi/2$ gave us the lowest matches
of any angles between $0$ and $\pi/2$ and shows that 
the minimax match of the $\Delta_{fine}$ and $\Delta_{med}$ resolutions
is $\Mminimax \ge 0.97$.  While the match of the coarsest resolution run at higher masses
is $\Mminimax \sim 0.92$, which indicates that the coarse resolution 
is not good enough for meeting the threshold when $a=0.8$ and $\theta\sim \pi/2$. 
For the rest of the paper, we will use the finest resolution for all spin cases
to ensure that the errors due to resolution are no greater $3\%$ in the most
difficult case to resolve: the high spin, high angle case, and should be much lesser in other cases.

Fig.~\ref{fig:spin0.8reslns} shows two features of the minimax match, namely it decreases
with increasing $\theta$ and with increasing $m$. 
The decrease of the match in Fig.~\ref{fig:spin0.8reslns} at higher mass is in contrast
to Fig.~\ref{fig:Mvsmass}.  There are two differences between the figures,
one is the phase optimization which, from Tab.~\ref{tab:optim}, we can see does not 
change the trend of the match versus mass.  The second difference is the
inclusion of higher modes.  The overall match will be reduced with non-zero $\theta$
since the frequencies of the modes increase with $\ell$, placing stronger resolution requirements.
The highest spin we analyze, $a=0.8$, has the lowest minimax matches 
of all the spin cases, as low as $\Mminimax = 0.92$ at the largest angle.
Large spins demand more resolution in general and this is seen particularly when resolving
modes with $\ell > 2$.  For the lower spin runs, $a \le 0.6$, the waveforms
are well resolved and the matches are all $\ge 0.99$ over the entire 
mass range even for $\theta \sim \pi/2$.  The decrease in match is seen most strongly at the larger mass range
since this mass range targets the merger and ringdown part of the signal.
We expect these match values and all the matches reported in this section 
to change for waveforms with more gravitational wave cycles since
the numerical errors will grow as the length of the run increases and
the accumulating phase errors will likely play a larger role.

\subsection{The Variation of Spinning Templates with Inclination Angle}
\label{sec:inclination}
An observer on the orbital axis of a merging binary, at  $\theta=\phi=0$,
will see a fully circularly
polarized waveform where $\ell=|m|=2$ is the only nonzero mode.
An observer, however, in the orbital plane of the binary or the equatorial
plane of the final black hole will see linearly polarized radiation
with the contribution from $\ell=|m|=2$ at its minimum.  
All other observers with intermediate orientations
would see elliptically polarized waves that are combinations of all the modes.  
In practice, numerical relativity can only resolve a finite set of modes.
Since the orientation of any given binary will be unknown {\it a. priori},      
we analyze the templates' dependence on the inclination and azimuthal angles.
The location of the detector in the center of mass frame fixes the mode content and the
orientation of the detector fixes the initial phase of the
waveform. 

\begin{figure}[h]
\includegraphics[height=7cm]{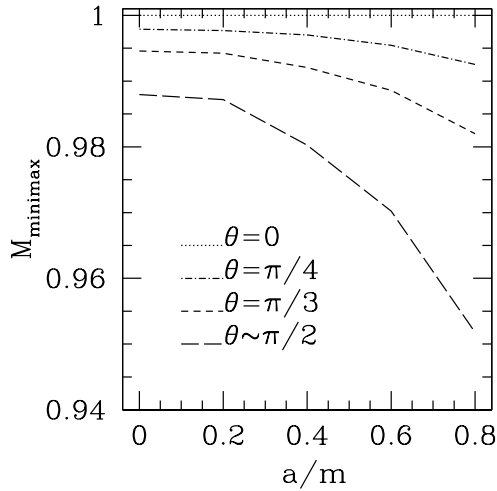}
\caption{The minimax match, 
$\Mminimax[h_1(\theta=0)],h_2(\theta_i)]$, 
vs $a$ is plotted for the values $\theta_i=\{0,\pi/4,\pi/3,\pi/2\}$ 
for the case $m=100M_\odot$.}
\label{fig:mmvsa}
\end{figure}

The resolution will now be fixed to the finest for the spin cases 
and the corresponding resolution for the non-spinning case. 
We use the minimax match as defined in Eq.~(\ref{eqn:minimax}), to see how
different the $\ell=|m|=2$ waveforms, given by $h(\theta=0,\phi=0)$,
are from the full waveforms, $h(\theta,\phi)$, for a given $a$. 
Our focus is on $\theta$ since it causes
a larger variation in the match than $\phi$.  To reduce the number
of parameters considered, and we set $\phi=\pi/2$ for the rest of
the paper when $\theta \neq 0$.  If we were to relax this condition,
the minimax matches would vary on the order of a percent. 
To observe the
variation with $\theta$, we first fix the total mass for both 
templates, $h_1$ and $h_2$, to $100M_\odot$.
In Fig.~\ref{fig:mmvsa} we present the minimax matches as a function of $a$
for templates at $\theta=\{0,\pi/4,\pi/3,\pi/2\}$.
By holding the target template to $\theta=0$ for each spin case, we can determine
at what angle and spin the matches drop below the threshold. 
In the case of $\theta=0$, the templates have the same parameters
and $\Mminimax = 1$ as it should. 
For $a=0$, all the templates match the $\theta=0$ target within $\Mminimax = 0.98$.
The plot also indicates that $\theta=0$ is close to the full waveform for
$\theta \le \pi/3$ for all of the spin cases.  For $\theta>\pi/3$, however,
the match drops below 0.98 for $a>0.2$.    
As the inclination with the
axis increases, the higher modes become important with
increasing spin. 

\begin{figure}[h]
\includegraphics[height=7cm]{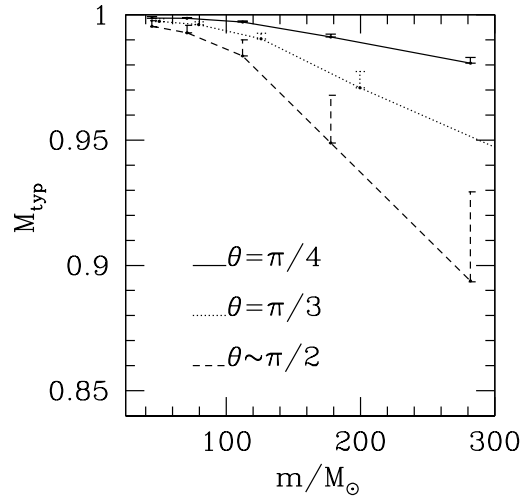}
\caption{The range of phase optimized matches between the $\theta=0$ waveform and the
full waveform for a low spin case, $a=0.2$.  The curves denote typical match 
$\Mtyp[h_1(\theta=0),h_{2+}(\theta_i)]$ for 
$\theta_i =\{\pi/4, \pi/3,\sim \pi/2\}$.   The lower end of the error bar is given by the
minimax match and the higher end is given by the best match.   The phase
optimization here is done over the phase of the $h_1$ template.}
\label{fig:h22hfulla0.2}
\end{figure}

\begin{figure}[h]
\includegraphics[height=7cm]{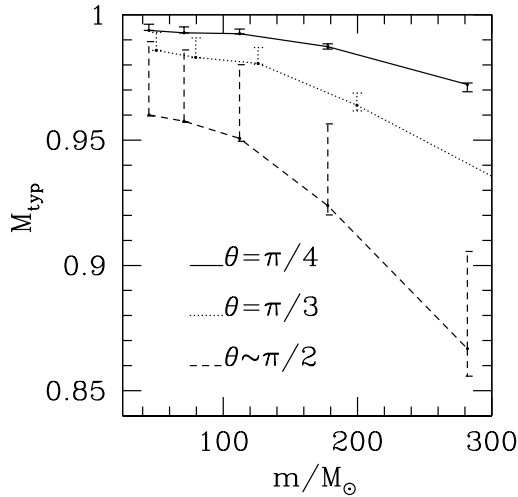}
\caption{The range of phase optimized matches between the $\theta=0$ waveform and the
full waveform for a high spin case, $a=0.8$.  The curves denote typical match 
$\Mtyp[h_1(\theta=0),h_{2+}(\theta_i)]$ for 
$\theta_i =\{\pi/4, \pi/3,\sim \pi/2\}$.   The lower end of the error bar is given by the
minimax match and the higher end is given by the best match.   The phase
optimization here is done over the phase of the $h_1$ template.}
\label{fig:h22hfulla0.8}
\end{figure}

To study inclination variation as a function
of mass,  we add the scale with total mass coming from the LIGO
noise curve.  We limit the presentation to two spins, $a=0.2$ and $a=0.8$
for clarity.   The best, minimax and typical matches are plotted
as a function of $m$ in Fig.~\ref{fig:h22hfulla0.2}
and Fig.~\ref{fig:h22hfulla0.8}.  The matches are between 
$h_1(\theta = 0)$ and
$h_2(\theta_i)$ for  three inclination angles  each
$\theta_i = \pi/4, \pi/3, \sim \pi/2$. 
We plot all three matches to demonstrate the range that the matches
can take depending on the choice of phase optimization.
The typical match is given by the lines and the
minimax and best matches are specified as the lower and upper error bars respectively.
One can see that the minimax match 
between the two templates sets a lower bound on the phase optimized 
matches.  The minimax match dips below 0.98 for $\theta>\pi/3$ and 
$m>100_\odot$ for both spin cases;  
and, unfortunately, dips below 0.98 for the
entire mass range for $\theta \sim \pi/2$ when $a=0.8$. 

An even more stringent test is the best match.  If the best match of a given
angle waveform is smaller than
some threshold then we would know that the higher modes are
significant and need to be used in creating the template bank.
We note that for both spin cases the best match is below 0.98 for
$\theta \sim \pi/2$ at masses greater than $170 M_\odot$ indicating
the need to include higher modes to make a detection
at large inclination and high masses.

One can see from the plots that the lower bound of the match between the circularly
polarized and the highly elliptically polarized
waveform is lower in the high mass cases. The decrease
in the  match at larger total masses may be indicating that the ringdown 
is more sensitive to the
presence of higher modes than the merger itself, since the deviation from 0.98 is larger
than expected from numerical error alone. 
These figures also suggest that  
the $\ell=|m|=2$ mode is closer to the full waveform at lower masses and
lower spins.  The fraction of events lost by not including
the higher modes needs to be calculated to predict the full impact.

The waveforms that we are using have been extracted from the numerical solution
of the Einstein equation at a finite radius.
In order to assess how the extraction
radius, $r_{\mathrm{ext}}$, affects the waveforms at a large extraction radius, we tabulate
the $\Mminimax$ of the $a=0.8$ waveforms in Tab.~\ref{tab:Randphi}. 
The matches are computed for a template 
fixed at $\theta=0$ and the target template at four values of $\theta$
to measure how much of the variation with $\theta$ depends on $r_{\mathrm{ext}}$.
The variation with the extraction radius is of the order of $\sim 1.3 \%$ 
in the worst case.  There is an interplay between better dynamical quality
and increased numerical error as $r_{\mathrm{ext}}$ is increased, which
is more noticeable in modes higher than $\ell =2$. For the rest of the 
paper we use the waveforms at $r_{\mathrm{ext}}=30m$.
\begin{table}[h]
\caption{Minimax matches for $m=100M_\odot$ and $a=0.8$
between two waveforms at the same extraction radius but different inclination
angle given by $\Mminimax[h_1(\theta=0),h_2(\theta_2)]\,.$ }
\begin{center}
\begin{tabular}{c c c c c }
\hline
 $r_{\mathrm{ext}} $ & $\theta_2=0$ & $\theta_2=\pi/4$ & $\theta_2=\pi/3$ & $\theta_2 \sim \pi/2$ \\[1mm]
\hline 
\hline
$30$ & $0.9779 $ & $ 0.9602$ & $0.9538$ & $0.9518$ \\[1mm] 
$40$ & $0.9773 $ & $ 0.9530$ & $0.9465$ & $0.9519$ \\[1mm]
$50$ & $0.9756 $ & $ 0.9473$ & $0.9423$ & $0.9546$ \\[1mm] 
$60$ & $0.9734 $ & $ 0.9445$ & $0.9409$ & $0.9564$ \\[1mm]
 \hline
\end{tabular}
\end{center}
\label{tab:Randphi}
\end{table}      

\subsection{Comparing the Various Spin Configuration Waveforms}
\label{sec:comparisons}
Given the computational cost of producing templates for \bbh{s},
we investigate how well a reduced template matches a target template
of arbitrary spin and inclination angle at a fixed resolution
($\Delta = m/40$).
We will study several cases, 
including spin configurations relative to each other at different
inclination angles for a fixed mass of $100 M\odot$
and then we specialize to a few angles and study the typical, minimax and the
best phase optimized matches. 

The simplest approach to compare the spinning waveforms
is to fix $\theta=0$ for both the target and template.
We then choose the template to have some spin and calculate
the match of that template with a target that varies with $a$, such
that we calculate $\Mminimax[h_1(a_i),h_2(a_j)]$ where
$a_i$ and $a_j$ run over all the combinations of spin. 
The minimax matches between all combinations of the spins are $\ge0.995$
over the mass range considered. This reflects the fact that the binaries
approach the same final black hole and that the initial spin gets radiated
away in modes other than $\ell=|m|=2$, where they are almost
identical during the late stages of the binary merger (sometimes called
universality \cite{NASA2006BBH}).  
This would mean that we could not distinguish between the non-spinning
and spinning waveforms making the  non-spinning
case sufficient for a detection template bank
 but potentially making parameter estimation problematic.
This degeneracy with spin, however, does not
hold when we include more radiation modes.

\begin{figure}[h]
\includegraphics[height=7cm]{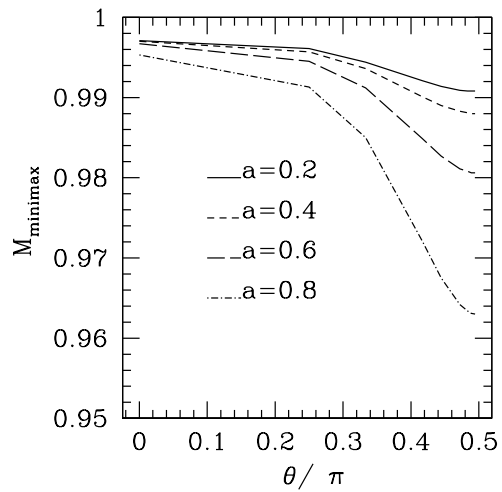}
\caption{The minimax match $\Mminimax[
h_1(a=0),h_2(a_i)]$ as a function of
$\theta$ between the waveforms of different 
spins at the same resolution $m/40$, with the non-spinning waveform at
resolution $m/38.4$. The mass of the final black hole for all the
cases is $100 M_\odot$ and the spins run over $a_i=\{0.2, 0.4, 0.6, 0.8\}$.
Note the monotonic decrease of the match with angle. }
\label{fig:angleminmaxspin0}
\end{figure}

\begin{figure}[h]
\includegraphics[height=7cm]{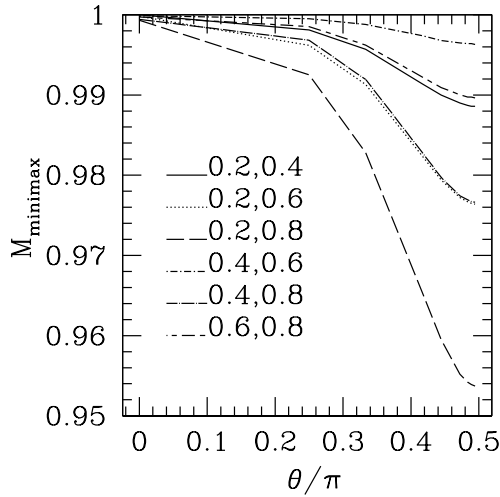}
\caption{The minimax match between the waveforms of different nonzero
initial spins, $ \Mminimax[
h_1(a_i),h_2(a_j)]$ as a function
of $\theta$. The mass of the final black hole for all the
cases is $100 M_\odot$ and $\{a_i , a_j\}$ runs over $\{0.2, 0.4, 0.6,
0.8\}$.  Note the monotonic decrease
of the minimax match with the angle $\theta$. }
\label{fig:angleminmaxallspins}
\end{figure}

We explore the matches between different
spin templates at different inclination angles, considering 
first the templates with a fixed mass of $100 M_\odot$.
The variation of the minimax match between a template of $a=0$ and 
target templates that vary with $a$ is presented in Fig.~\ref{fig:angleminmaxspin0}  
and with both the template and target varying with $a \ne 0$ 
in Fig.~\ref{fig:angleminmaxallspins}.  In terms of the match,
Fig.~\ref{fig:angleminmaxspin0} corresponds to a minimax match, 
$\Mminimax[h_1(a=0),h_2(a_i)]$, 
and Fig.~\ref{fig:angleminmaxallspins} between six combinations of the four
spinning configurations, $\Mminimax[h_1(a_i),h_2(a_j)]$ both versus $\theta$.
In both figures, at $\theta \leq \pi/3$ the template is within
a $\Mminimax$ of 0.98 with the targets, making them indistinguishable.
When the template is fixed at $a=0$ in Fig.~\ref{fig:angleminmaxspin0}, 
the match is $>0.98$ for all $\theta$ except with $a>0.6$, which shows a stronger
drop-off with angle for higher initial spins.  
Fig.~\ref{fig:angleminmaxallspins} shows  that the matches between adjacent
templates among the $a=\{0.2,0.4,0.6,0.8\}$ group are better than $~0.99$, while
the matches that have a spin parameter difference 
$a_1 - a_2 \ge 0.4$ drop below $0.98$ at higher angles. These
templates are based on short waveforms of just two and a half cycles and
the matches will likely get worse when more cycles are included. 
We conclude from these plots that the $\ell=|m|=2$ mode continues to 
dominate the other modes at small angles, $\theta<\pi/4$, leading
to  similar matches between a spinning and non-spinning configuration.  
As the inclination angle grows, however, the
presence of higher modes becomes more pronounced, breaking the degeneracy
between the waveforms from different initial spin configurations.
 
For completeness, we present the table of the range of values
over which the phase optimized matches of the different spinning
waveforms can vary in Tab.~\ref{tab:range}. These matches are evaluated at 
$\theta \sim \pi/2$ to report the widest range of matches calculated.
The table indicates that the range of matches is higher at lower masses
and tends to decrease with mass.
\begin{table}[h]
\caption{ The range of phase optimized matches reported as $\left(\Mminimax,\Mbest\right)$
between two templates of various spin pairs $\left(a_1,a_2\right)$  where both templates
have $\theta \sim \pi/2.$}
\begin{center}
\begin{tabular}{c c c c }
\hline
$(a_1,a_2)$ &$ 50 M_\odot$&$  120 M_\odot$&$300 M_\odot$\\
\hline
\hline
$\left(0.0,0.2 \right)$& $ \left(0.8940, 0.9513\right)$&$ \left(0.9228, 0.9618\right)$&$ \left(0.9446, 0.9701\right)$ \\ [1 mm]
$ \left(0.0,0.4 \right)$&$  \left(0.8570, 0.9507\right)$&$ \left(0.8882, 0.9604\right)$&$ \left(0.9141, 0.9684\right)$\\ [1 mm]
$\left(0.0,0.6 \right)$& $ \left(0.8595, 0.9530\right)$&$ \left(0.8898, 0.9618\right)$&$ \left(0.8234, 0.9427\right)$\\ [1 mm]
 $ \left(0.0,0.8 \right)$&  $\left(0.7325, 0.9196\right)$&$ \left(0.7896, 0.9326\right)$&$ \left(0.8369, 0.9449\right)$\\ [1 mm]
 $ \left(0.2,0.4 \right)$&$\left(0.9860, 0.9964\right)$&$ \left( 0.9884, 0.9967\right)$&$ \left(0.9858, 0.9954\right)$\\ [1 mm]
 $ \left(0.2,0.6\right)$&$\left(0.9742, 0.9927\right)$&$ \left( 0.9751, 0.9927,\right)$&$ \left( 0.9665, 0.9912\right)$\\ [1 mm]
$ \left(0.2,0.8\right)$&$\left(0.9526, 0.9832\right)$&$ \left( 0.9497, 0.9793,\right)$&$ \left(  0.9194, 0.9681\right)$\\ [1 mm]
 $ \left(0.4,0.6\right)$&$\left(0.9973, 0.9993\right)$&$ \left(0.9959, 0.9988, \right)$&$\left(0.9945, 0.9971\right)$\\ [1 mm]
$ \left(0.4,0.8\right)$& $\left(0.9808, 0.9956\right)$&$ \left(0.9732, 0.9902,\right)$&$ \left( 0.9611, 0.9738\right)$\\ [1 mm]
$ \left(0.6,0.8\right)$&$\left(0.9911, 0.9982\right)$&$ \left( 0.9880, 0.9950, \right)$&$ \left(0.9824, 0.9867\right)$\\ [1 mm]
 \hline
 \hline
\end{tabular}
\end{center}
\label{tab:range}
\end{table}   

\begin{figure}[h]
\includegraphics[height=7cm]{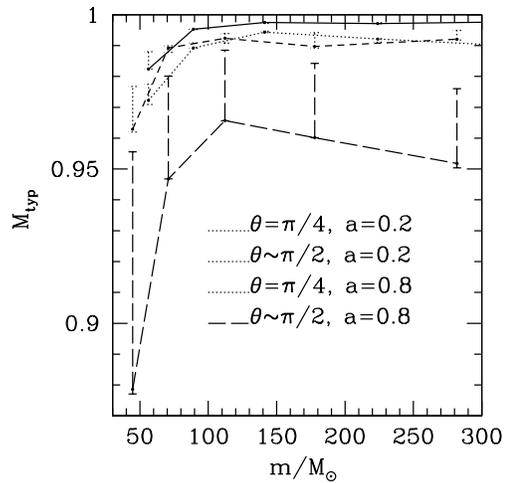}
\caption{We plot the range and the typical values of phase optimized match between the
waveforms of different initial spins, 
$\Mtyp[h_{+1}(a=0,\theta_j),h_2(a_i,\theta_j)]$ as a function
of mass where $a_i =\{0.2, 0.8\}$ and $\theta_j=\{\pi/4,\pi/2\}$. }
\label{fig:typmatcha0.2a0.8}
\end{figure}

To place the minimax matches between the spins 
in context with the typical and the best matches, 
we present the typical match $\Mtyp[h_{+1}(a=0),h_2(a=a_j)]$ 
for $a_j=\{0.2, 0.8\}$ and
$\theta= \{\pi/2, \pi/4\}$ in Fig.~\ref{fig:typmatcha0.2a0.8}.
The error bars are  given by the minimax and the best match in order
to demonstrate their  variation with mass. 
As seen in Fig.~\ref{fig:typmatcha0.2a0.8}, 
the difference between the minimax and best matches is slight 
at small values of $\theta$ and $a$,
while for large angles it can vary over the order of $\sim 3\%$.
The variation is much higher at lower masses because 
there are more cycles in the sensitive part of the LIGO noise curve
making the match more sensitive to the relative phasing of the templates.
The figure indicates that using a non-spinning waveform as the template
in detection would cause the largest losses 
when the target has large spin and angle.

We have seen the variation of the match between templates of
the same spin parameters and different angles and different spin 
parameters and the same angle. In setting up a
template bank using these numerical waveforms,
it may be useful to know if  just a couple of templates at some specific
inclination and spins can cover the whole set of waveforms
considered here.
 
\begin{figure}[h]
\includegraphics[height=7cm]{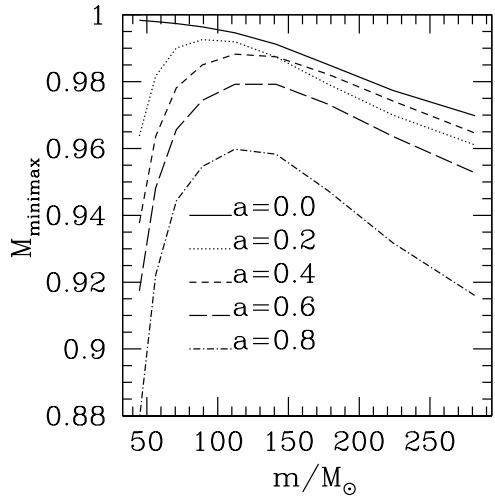}
\caption{We plot the minimax match between the
target templates of different initial spins at a fixed $\theta$
with a template of $a=0.0$, 
$\Mminimax[h_1(a=0,\theta=\pi/4),h_2(a_i,\theta=\pi/2)]$ 
as a function of mass. Notice the low matches at the smaller masses.}
\label{fig:matchminfixedtheta0.0}
\end{figure}

\begin{figure}[h]
\includegraphics[height=7cm]{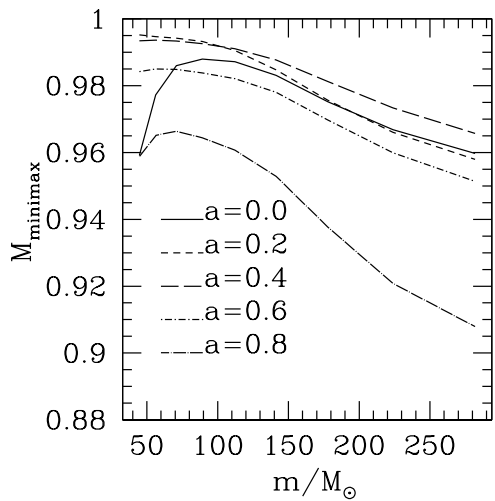}
\caption{We plot the minimax match between the
target templates of different initial spins at a fixed 
$\theta$ with a template of $a=0.4$, 
$\Mminimax[h_1(a=0.4,\theta=\pi/4),h_2(a_i,\theta=\pi/2)]$
as a function of mass.}
\label{fig:matchminfixedtheta0.4}
\end{figure}

For this study we calculate the minimax matches between a template 
of a given value of spin and inclination angle and a target with a 
spin parameter varying over $a_i =\{0.0,0.2,0.4,0.6,0.8\}$
and a fixed, but different inclination angle.
In Fig.~\ref{fig:matchminfixedtheta0.0} we plot the variation of the minimax match 
for a template with $a_1=0.0$ and $\theta_1 = \pi/4$ with 
target templates of varying spin 
and $\theta_2 \sim \pi/2$ as a function of total mass.  
We made the choice of the template to have $\theta_1=\pi/4$ since we
have concluded that a choice of $\theta=0$ is
going to fail to match well over a range of target inclination angles, especially  at
lower masses.  We choose the target template to have
$\theta_2\sim\pi/2$ to calculate the worse case scenario for the inclination angle
of the target.  In Fig.~\ref{fig:matchminfixedtheta0.4}  we
substitute a template of $a_1=0.4$ and keep all the other parameters
the same as in Fig.~\ref{fig:matchminfixedtheta0.0}.  This improves
the matches at smaller total mass.
One can see from the figures that the higher the match threshold, 
the smaller  the  mass range covered by the template.
A coarse template bank would do better
with $a=0.4$ and $\theta=\pi/4$ than $a=0$ at the same $\theta=\pi/4$
especially for the lower masses and hints that there
may be an optimal way to lay out this template space.  
Another feature is that for higher masses, the matches
for both the cases look similar, as the late-merger/ringdown 
has some universality-independence from the initial spin.

The use of these axial-spin waveforms as templates for detection has some limitations.
The waveforms investigated had the angle between the initial spins kept fixed,
and it is not clear how the conclusions we draw from this study hold as this condition
is relaxed.  Although it is likely that the higher radiation modes will
play an even bigger role than they do for the relatively simple case investigated here.
The major limitation however, is the small
number of gravitational wave cycles present in the waveforms.
The ISCO frequency is about $0.02/m$.  The mass at which the 
ISCO frequency hits the seismic wall of LIGO noise curve of $40\,\Hz$
is about $100 M_\odot$.  If one allows the 
lower limit in the match integrals to vary or be fixed at 
$0.02/m=70\,\Hz$ for example, then we can trust these matches up to 
$\approx 60M_\odot$  
assuming that we will search
for events that are merging in the LIGO band with little to no inspiral.
Although the results  will not change too much at
high masses even if inspiral is included in the numerical templates, 
low mass results are expected to have lower matches.
Finally, another limitation of these matches is that it 
assumes identical total mass for the template and the waveform being 
compared to it. This assumption needs to be relaxed to estimate the 
"effectualness" of templates for detection, which will be done in a future work.
  
\section{Conclusion}
\label{sec:conclusion}
The parameter space of the \bbh{} system is rapidly being explored by the
numerical relativity community.  Given how much time and effort in 
takes to evolve the \bbh{} spacetimes and then to use
those waveforms in detection schemes, it is becoming increasingly
important to probe the parameter space of the \bbh{} system as the waveforms
become available.  With this motivation, we have investigated a series
of waveforms from the final two to three orbits of a binary-black-hole
system made up of equal but spinning black holes with matched filtering.  
Although the waveforms we use are not long enough in terms of gravitational wave cycles 
to cover the entire LIGO frequency band, they shed light on what weight different parameters
will have in the matched filtering procedure using LIGO's initial noise curve. 
We study the dependence of the match on the intrinsic parameters of numerical
resolution, inclination and spin as well as the extrinsic parameters of
arrival time and phase.

Given the numerical convergence properties of our waveforms, we first predict and
show that the match has a dependence on the resolution of
$\Delta^{2p}$, in which a fourth order evolution scheme has $p=4$.  
For the resolutions of our \bbh{} evolutions, we get very high matches ($>0.98$) 
for resolutions above $m/32$ when $a=0$  and above $m/38$ for waveforms
with $a=0.8$.  The resolutions are only appropriate for the cases we have investigated.
For the waveforms generated out of the $\ell=|m|=2$ mode,
we find that the match values between the same waveforms at different resolution decreases
as the total mass decreases. 
We only report matches from $50M_\odot-300M_\odot$, where $50M_\odot$ is set
by our initial orbital velocity.

We then include the variation of the match with the inclination angle.  It is often convenient
and instructive to extract the gravitational waveform, $\Psi_4$, decomposed
into spin-weighted spherical harmonics as a function of $\ell$ and $m$.  Including modes higher
than the dominant mode, $\ell=|m|=2$, is equivalent to allowing the inclination angle
of the binary system to vary.  Only at $\theta=0$ would the $\ell=|m|=2$ mode
be the only harmonic in the waveform.  The target waveform will have an unknown orientation; 
and, therefore
we vary the inclination to study what effect this may have on detection.  
As we include all the modes in the analysis
the match decreases, to as low as $\approx 0.85$, must notably for larger masses and higher spins. 
Some decrease in match is expected since the higher modes require more resolution; 
however, the decrease was greater than that accounted for by the resolution alone.  

Finally, in light of the computational effort involved in searching over large, densely populated
template banks, we calculate the match between waveforms of different initial spin 
to see if a reduced set of spinning waveforms would be good enough.
The waveforms we use in this analysis evolve toward the same Kerr black hole ($a\approx 0.66)$
even though the initial spins of the individual black holes vary from $0.0$ to $0.8$ 
because of the fixed initial ADM angular momentum.  This results in very high matches,
$\Mminimax > 0.99$,
between all the spinning and non-spinning waveforms for the dominant mode.  Without
adding information from the higher radiation modes, all the spinning \bbh{}
waveforms appeared the same in terms of matched filtering.  

Once $h$ was constructed using modes $\ell<5$, this degeneracy broke.
We constructed two "typical'' cases, one in which the template had a 
fixed $\theta=\pi/2$ and $a=0$ and the other $a=0.4$.
The templates were matched against a target template also at $\theta=\pi/2$ 
for each spinning waveforms.
When $a=0$, the matches were as low as $0.88$ versus $m$, but were
$>0.95$ for $m>100M_\odot$.
By choosing an intermediate spin
of $a=0.4$, the matches were improved for the lower masses, $m<100M_\odot$, increasing
the lowest value from 0.88 to 0.96. We speculate that 
this indicates that there may be an optimal layout for the template space for a coarse search  
over spins.
The matches were always at their highest at $100M_\odot$, although not necessarily
greater than 0.98.

In summary, we have investigated the sensitivity of matched filtering to the presence
of spin and radiation modes in a limited case of spinning \bbh{} waveforms.  We find
that if we ignore the modes higher than $\ell=2$, there is very little difference between
the spinning waveforms, especially at large $m$ due to our choice of initial data.
We do find, however, that the inclusion of larger $\ell$ changes this picture in two
ways.  Two templates with the same spin but different inclination only match for
low values of the inclination angle and two templates of the same inclination angle
but different spin only match when the spins are within 0.2 of each other.
Matching well means that $\Mminimax \ge 0.98$.
In general, the higher modes reduce the matches more at large masses than
at small masses.  It remains to be seen 
how these results change when the spins are no longer fixed to axial
configurations, but more elaborate initial data will likely vary more with
$\ell$.   
 
\section{Acknowledgments}
We thank Shane Larson, Sam Finn, Pablo Laguna and Ben Owen for 
their insightful discussions on topics in this paper.
We thank The Center for Gravitational Wave Physics is supported by the
NSF under cooperative agreement PHY 01-14375.  Work partially
supported by NSF grant PHY-0354821 to DS.
 
\bibliographystyle{apsrev} \bibliography{paper}
\end{document}

%% file: amp-all-include.tex
\begingroup
  \makeatletter
  \providecommand\color[2][]{%
    \GenericError{(gnuplot) \space\space\space\@spaces}{%
      Package color not loaded in conjunction with
      terminal option `colourtext'%
    }{See the gnuplot documentation for explanation.%
    }{Either use 'blacktext' in gnuplot or load the package
      color.sty in LaTeX.}%
    \renewcommand\color[2][]{}%
  }%
  \providecommand\includegraphics[2][]{%
    \GenericError{(gnuplot) \space\space\space\@spaces}{%
      Package graphicx or graphics not loaded%
    }{See the gnuplot documentation for explanation.%
    }{The gnuplot epslatex terminal needs graphicx.sty or graphics.sty.}%
    \renewcommand\includegraphics[2][]{}%
  }%
  \providecommand\rotatebox[2]{#2}%
  \@ifundefined{ifGPcolor}{%
    \newif\ifGPcolor
    \GPcolorfalse
  }{}%
  \@ifundefined{ifGPblacktext}{%
    \newif\ifGPblacktext
    \GPblacktexttrue
  }{}%
  \let\gplgaddtomacro\g@addto@macro
  \gdef\gplbacktext{}%
  \gdef\gplfronttext{}%
  \makeatother
  \ifGPblacktext
    \def\colorrgb#1{}%
    \def\colorgray#1{}%
  \else
    \ifGPcolor
      \def\colorrgb#1{\color[rgb]{#1}}%
      \def\colorgray#1{\color[gray]{#1}}%
      \expandafter\def\csname LTw\endcsname{\color{white}}%
      \expandafter\def\csname LTb\endcsname{\color{black}}%
      \expandafter\def\csname LTa\endcsname{\color{black}}%
      \expandafter\def\csname LT0\endcsname{\color[rgb]{1,0,0}}%
      \expandafter\def\csname LT1\endcsname{\color[rgb]{0,1,0}}%
      \expandafter\def\csname LT2\endcsname{\color[rgb]{0,0,1}}%
      \expandafter\def\csname LT3\endcsname{\color[rgb]{1,0,1}}%
      \expandafter\def\csname LT4\endcsname{\color[rgb]{0,1,1}}%
      \expandafter\def\csname LT5\endcsname{\color[rgb]{1,1,0}}%
      \expandafter\def\csname LT6\endcsname{\color[rgb]{0,0,0}}%
      \expandafter\def\csname LT7\endcsname{\color[rgb]{1,0.3,0}}%
      \expandafter\def\csname LT8\endcsname{\color[rgb]{0.5,0.5,0.5}}%
    \else
      \def\colorrgb#1{\color{black}}%
      \def\colorgray#1{\color[gray]{#1}}%
      \expandafter\def\csname LTw\endcsname{\color{white}}%
      \expandafter\def\csname LTb\endcsname{\color{black}}%
      \expandafter\def\csname LTa\endcsname{\color{black}}%
      \expandafter\def\csname LT0\endcsname{\color{black}}%
      \expandafter\def\csname LT1\endcsname{\color{black}}%
      \expandafter\def\csname LT2\endcsname{\color{black}}%
      \expandafter\def\csname LT3\endcsname{\color{black}}%
      \expandafter\def\csname LT4\endcsname{\color{black}}%
      \expandafter\def\csname LT5\endcsname{\color{black}}%
      \expandafter\def\csname LT6\endcsname{\color{black}}%
      \expandafter\def\csname LT7\endcsname{\color{black}}%
      \expandafter\def\csname LT8\endcsname{\color{black}}%
    \fi
  \fi
  \setlength{\unitlength}{0.0500bp}%
  \begin{picture}(5040.00,3528.00)%
    \gplgaddtomacro\gplbacktext{%
      \csname LTb\endcsname%
      \put(1166,660){\makebox(0,0)[r]{\strut{} 0.002}}%
      \put(1166,1094){\makebox(0,0)[r]{\strut{} 0.0021}}%
      \put(1166,1528){\makebox(0,0)[r]{\strut{} 0.0022}}%
      \put(1166,1962){\makebox(0,0)[r]{\strut{} 0.0023}}%
      \put(1166,2396){\makebox(0,0)[r]{\strut{} 0.0024}}%
      \put(1166,2830){\makebox(0,0)[r]{\strut{} 0.0025}}%
      \put(1166,3264){\makebox(0,0)[r]{\strut{} 0.0026}}%
      \put(1547,440){\makebox(0,0){\strut{} 195}}%
      \put(2171,440){\makebox(0,0){\strut{} 200}}%
      \put(2795,440){\makebox(0,0){\strut{} 205}}%
      \put(3419,440){\makebox(0,0){\strut{} 210}}%
      \put(4042,440){\makebox(0,0){\strut{} 215}}%
      \put(4666,440){\makebox(0,0){\strut{} 220}}%
      \put(396,1962){\rotatebox{90}{\makebox(0,0){\strut{}$|\Psi_4^{2,2}|$}}}%
      \put(2982,110){\makebox(0,0){\strut{}$t$}}%
    }%
    \gplgaddtomacro\gplfronttext{%
      \csname LTb\endcsname%
      \put(3679,3058){\makebox(0,0)[r]{\strut{}R1a}}%
      \csname LTb\endcsname%
      \put(3679,2772){\makebox(0,0)[r]{\strut{}R1b}}%
      \csname LTb\endcsname%
      \put(3679,2486){\makebox(0,0)[r]{\strut{}R1c}}%
      \csname LTb\endcsname%
      \put(3679,2200){\makebox(0,0)[r]{\strut{}R1d}}%
      \csname LTb\endcsname%
      \put(3679,1914){\makebox(0,0)[r]{\strut{}R1e}}%
      \csname LTb\endcsname%
      \put(3679,1628){\makebox(0,0)[r]{\strut{}R1f}}%
    }%
    \gplbacktext
    \put(0,0){\includegraphics{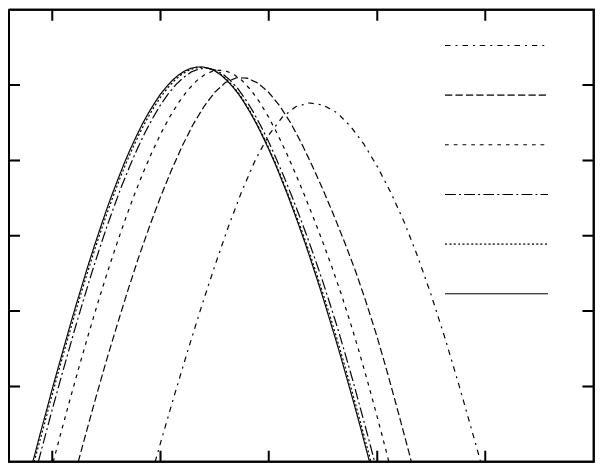}}%
    \gplfronttext
  \end{picture}%
\endgroup

%% file: convergence-re-include.tex
\begingroup
  \makeatletter
  \providecommand\color[2][]{%
    \GenericError{(gnuplot) \space\space\space\@spaces}{%
      Package color not loaded in conjunction with
      terminal option `colourtext'%
    }{See the gnuplot documentation for explanation.%
    }{Either use 'blacktext' in gnuplot or load the package
      color.sty in LaTeX.}%
    \renewcommand\color[2][]{}%
  }%
  \providecommand\includegraphics[2][]{%
    \GenericError{(gnuplot) \space\space\space\@spaces}{%
      Package graphicx or graphics not loaded%
    }{See the gnuplot documentation for explanation.%
    }{The gnuplot epslatex terminal needs graphicx.sty or graphics.sty.}%
    \renewcommand\includegraphics[2][]{}%
  }%
  \providecommand\rotatebox[2]{#2}%
  \@ifundefined{ifGPcolor}{%
    \newif\ifGPcolor
    \GPcolorfalse
  }{}%
  \@ifundefined{ifGPblacktext}{%
    \newif\ifGPblacktext
    \GPblacktexttrue
  }{}%
  \let\gplgaddtomacro\g@addto@macro
  \gdef\gplbacktext{}%
  \gdef\gplfronttext{}%
  \makeatother
  \ifGPblacktext
    \def\colorrgb#1{}%
    \def\colorgray#1{}%
  \else
    \ifGPcolor
      \def\colorrgb#1{\color[rgb]{#1}}%
      \def\colorgray#1{\color[gray]{#1}}%
      \expandafter\def\csname LTw\endcsname{\color{white}}%
      \expandafter\def\csname LTb\endcsname{\color{black}}%
      \expandafter\def\csname LTa\endcsname{\color{black}}%
      \expandafter\def\csname LT0\endcsname{\color[rgb]{1,0,0}}%
      \expandafter\def\csname LT1\endcsname{\color[rgb]{0,1,0}}%
      \expandafter\def\csname LT2\endcsname{\color[rgb]{0,0,1}}%
      \expandafter\def\csname LT3\endcsname{\color[rgb]{1,0,1}}%
      \expandafter\def\csname LT4\endcsname{\color[rgb]{0,1,1}}%
      \expandafter\def\csname LT5\endcsname{\color[rgb]{1,1,0}}%
      \expandafter\def\csname LT6\endcsname{\color[rgb]{0,0,0}}%
      \expandafter\def\csname LT7\endcsname{\color[rgb]{1,0.3,0}}%
      \expandafter\def\csname LT8\endcsname{\color[rgb]{0.5,0.5,0.5}}%
    \else
      \def\colorrgb#1{\color{black}}%
      \def\colorgray#1{\color[gray]{#1}}%
      \expandafter\def\csname LTw\endcsname{\color{white}}%
      \expandafter\def\csname LTb\endcsname{\color{black}}%
      \expandafter\def\csname LTa\endcsname{\color{black}}%
      \expandafter\def\csname LT0\endcsname{\color{black}}%
      \expandafter\def\csname LT1\endcsname{\color{black}}%
      \expandafter\def\csname LT2\endcsname{\color{black}}%
      \expandafter\def\csname LT3\endcsname{\color{black}}%
      \expandafter\def\csname LT4\endcsname{\color{black}}%
      \expandafter\def\csname LT5\endcsname{\color{black}}%
      \expandafter\def\csname LT6\endcsname{\color{black}}%
      \expandafter\def\csname LT7\endcsname{\color{black}}%
      \expandafter\def\csname LT8\endcsname{\color{black}}%
    \fi
  \fi
  \setlength{\unitlength}{0.0500bp}%
  \begin{picture}(5040.00,3528.00)%
    \gplgaddtomacro\gplbacktext{%
      \csname LTb\endcsname%
      \put(1188,660){\makebox(0,0)[r]{\strut{}-0.003}}%
      \put(1188,1094){\makebox(0,0)[r]{\strut{}-0.002}}%
      \put(1188,1528){\makebox(0,0)[r]{\strut{}-0.001}}%
      \put(1188,1962){\makebox(0,0)[r]{\strut{}0.000}}%
      \put(1188,2396){\makebox(0,0)[r]{\strut{}0.001}}%
      \put(1188,2830){\makebox(0,0)[r]{\strut{}0.002}}%
      \put(1188,3264){\makebox(0,0)[r]{\strut{}0.003}}%
      \put(1320,440){\makebox(0,0){\strut{} 0}}%
      \put(1878,440){\makebox(0,0){\strut{} 50}}%
      \put(2435,440){\makebox(0,0){\strut{} 100}}%
      \put(2993,440){\makebox(0,0){\strut{} 150}}%
      \put(3551,440){\makebox(0,0){\strut{} 200}}%
      \put(4108,440){\makebox(0,0){\strut{} 250}}%
      \put(4666,440){\makebox(0,0){\strut{} 300}}%
      \put(418,1962){\rotatebox{90}{\makebox(0,0){\strut{}$\textrm{Re}\left [\Psi_4^{2,2}\right]$}}}%
      \put(2993,110){\makebox(0,0){\strut{}$t$}}%
    }%
    \gplgaddtomacro\gplfronttext{%
      \csname LTb\endcsname%
      \put(2307,3058){\makebox(0,0)[l]{\strut{}R1a}}%
      \csname LTb\endcsname%
      \put(2307,2772){\makebox(0,0)[l]{\strut{}R1b}}%
      \csname LTb\endcsname%
      \put(2307,2486){\makebox(0,0)[l]{\strut{}R1f}}%
    }%
    \gplbacktext
    \put(0,0){\includegraphics{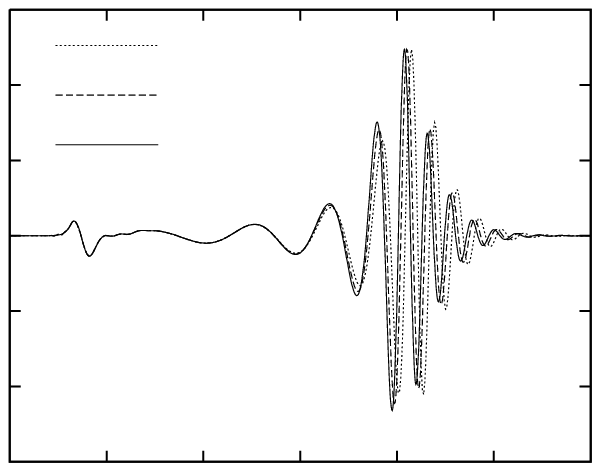}}%
    \gplfronttext
  \end{picture}%
\endgroup

%% file: convergence-amp-include.tex
\begingroup
  \makeatletter
  \providecommand\color[2][]{%
    \GenericError{(gnuplot) \space\space\space\@spaces}{%
      Package color not loaded in conjunction with
      terminal option `colourtext'%
    }{See the gnuplot documentation for explanation.%
    }{Either use 'blacktext' in gnuplot or load the package
      color.sty in LaTeX.}%
    \renewcommand\color[2][]{}%
  }%
  \providecommand\includegraphics[2][]{%
    \GenericError{(gnuplot) \space\space\space\@spaces}{%
      Package graphicx or graphics not loaded%
    }{See the gnuplot documentation for explanation.%
    }{The gnuplot epslatex terminal needs graphicx.sty or graphics.sty.}%
    \renewcommand\includegraphics[2][]{}%
  }%
  \providecommand\rotatebox[2]{#2}%
  \@ifundefined{ifGPcolor}{%
    \newif\ifGPcolor
    \GPcolorfalse
  }{}%
  \@ifundefined{ifGPblacktext}{%
    \newif\ifGPblacktext
    \GPblacktexttrue
  }{}%
  \let\gplgaddtomacro\g@addto@macro
  \gdef\gplbacktext{}%
  \gdef\gplfronttext{}%
  \makeatother
  \ifGPblacktext
    \def\colorrgb#1{}%
    \def\colorgray#1{}%
  \else
    \ifGPcolor
      \def\colorrgb#1{\color[rgb]{#1}}%
      \def\colorgray#1{\color[gray]{#1}}%
      \expandafter\def\csname LTw\endcsname{\color{white}}%
      \expandafter\def\csname LTb\endcsname{\color{black}}%
      \expandafter\def\csname LTa\endcsname{\color{black}}%
      \expandafter\def\csname LT0\endcsname{\color[rgb]{1,0,0}}%
      \expandafter\def\csname LT1\endcsname{\color[rgb]{0,1,0}}%
      \expandafter\def\csname LT2\endcsname{\color[rgb]{0,0,1}}%
      \expandafter\def\csname LT3\endcsname{\color[rgb]{1,0,1}}%
      \expandafter\def\csname LT4\endcsname{\color[rgb]{0,1,1}}%
      \expandafter\def\csname LT5\endcsname{\color[rgb]{1,1,0}}%
      \expandafter\def\csname LT6\endcsname{\color[rgb]{0,0,0}}%
      \expandafter\def\csname LT7\endcsname{\color[rgb]{1,0.3,0}}%
      \expandafter\def\csname LT8\endcsname{\color[rgb]{0.5,0.5,0.5}}%
    \else
      \def\colorrgb#1{\color{black}}%
      \def\colorgray#1{\color[gray]{#1}}%
      \expandafter\def\csname LTw\endcsname{\color{white}}%
      \expandafter\def\csname LTb\endcsname{\color{black}}%
      \expandafter\def\csname LTa\endcsname{\color{black}}%
      \expandafter\def\csname LT0\endcsname{\color{black}}%
      \expandafter\def\csname LT1\endcsname{\color{black}}%
      \expandafter\def\csname LT2\endcsname{\color{black}}%
      \expandafter\def\csname LT3\endcsname{\color{black}}%
      \expandafter\def\csname LT4\endcsname{\color{black}}%
      \expandafter\def\csname LT5\endcsname{\color{black}}%
      \expandafter\def\csname LT6\endcsname{\color{black}}%
      \expandafter\def\csname LT7\endcsname{\color{black}}%
      \expandafter\def\csname LT8\endcsname{\color{black}}%
    \fi
  \fi
  \setlength{\unitlength}{0.0500bp}%
  \begin{picture}(5040.00,3528.00)%
    \gplgaddtomacro\gplbacktext{%
      \csname LTb\endcsname%
      \put(1188,897){\makebox(0,0)[r]{\strut{}-0.0004}}%
      \put(1188,1370){\makebox(0,0)[r]{\strut{}-0.0002}}%
      \put(1188,1844){\makebox(0,0)[r]{\strut{}0.0000}}%
      \put(1188,2317){\makebox(0,0)[r]{\strut{}0.0002}}%
      \put(1188,2791){\makebox(0,0)[r]{\strut{}0.0004}}%
      \put(1188,3264){\makebox(0,0)[r]{\strut{}0.0006}}%
      \put(1320,440){\makebox(0,0){\strut{} 0}}%
      \put(1878,440){\makebox(0,0){\strut{} 50}}%
      \put(2435,440){\makebox(0,0){\strut{} 100}}%
      \put(2993,440){\makebox(0,0){\strut{} 150}}%
      \put(3551,440){\makebox(0,0){\strut{} 200}}%
      \put(4108,440){\makebox(0,0){\strut{} 250}}%
      \put(4666,440){\makebox(0,0){\strut{} 300}}%
      \put(286,1962){\rotatebox{90}{\makebox(0,0){\strut{}$\delta |\Psi_4^{2,2}|$}}}%
      \put(2993,110){\makebox(0,0){\strut{}$t$}}%
    }%
    \gplgaddtomacro\gplfronttext{%
      \csname LTb\endcsname%
      \put(2307,3058){\makebox(0,0)[l]{\strut{}a - b}}%
      \csname LTb\endcsname%
      \put(2307,2772){\makebox(0,0)[l]{\strut{}$C_3 \times (b-f)$}}%
      \csname LTb\endcsname%
      \put(2307,2486){\makebox(0,0)[l]{\strut{}$C_4 \times (b-f)$}}%
      \csname LTb\endcsname%
      \put(2307,2200){\makebox(0,0)[l]{\strut{}$C_5 \times (b-f)$}}%
    }%
    \gplbacktext
    \put(0,0){\includegraphics{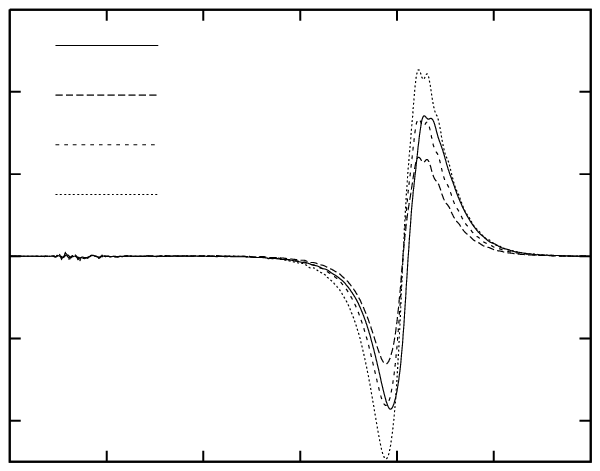}}%
    \gplfronttext
  \end{picture}%
\endgroup

%% file: convergence-phase-include.tex
\begingroup
  \makeatletter
  \providecommand\color[2][]{%
    \GenericError{(gnuplot) \space\space\space\@spaces}{%
      Package color not loaded in conjunction with
      terminal option `colourtext'%
    }{See the gnuplot documentation for explanation.%
    }{Either use 'blacktext' in gnuplot or load the package
      color.sty in LaTeX.}%
    \renewcommand\color[2][]{}%
  }%
  \providecommand\includegraphics[2][]{%
    \GenericError{(gnuplot) \space\space\space\@spaces}{%
      Package graphicx or graphics not loaded%
    }{See the gnuplot documentation for explanation.%
    }{The gnuplot epslatex terminal needs graphicx.sty or graphics.sty.}%
    \renewcommand\includegraphics[2][]{}%
  }%
  \providecommand\rotatebox[2]{#2}%
  \@ifundefined{ifGPcolor}{%
    \newif\ifGPcolor
    \GPcolorfalse
  }{}%
  \@ifundefined{ifGPblacktext}{%
    \newif\ifGPblacktext
    \GPblacktexttrue
  }{}%
  \let\gplgaddtomacro\g@addto@macro
  \gdef\gplbacktext{}%
  \gdef\gplfronttext{}%
  \makeatother
  \ifGPblacktext
    \def\colorrgb#1{}%
    \def\colorgray#1{}%
  \else
    \ifGPcolor
      \def\colorrgb#1{\color[rgb]{#1}}%
      \def\colorgray#1{\color[gray]{#1}}%
      \expandafter\def\csname LTw\endcsname{\color{white}}%
      \expandafter\def\csname LTb\endcsname{\color{black}}%
      \expandafter\def\csname LTa\endcsname{\color{black}}%
      \expandafter\def\csname LT0\endcsname{\color[rgb]{1,0,0}}%
      \expandafter\def\csname LT1\endcsname{\color[rgb]{0,1,0}}%
      \expandafter\def\csname LT2\endcsname{\color[rgb]{0,0,1}}%
      \expandafter\def\csname LT3\endcsname{\color[rgb]{1,0,1}}%
      \expandafter\def\csname LT4\endcsname{\color[rgb]{0,1,1}}%
      \expandafter\def\csname LT5\endcsname{\color[rgb]{1,1,0}}%
      \expandafter\def\csname LT6\endcsname{\color[rgb]{0,0,0}}%
      \expandafter\def\csname LT7\endcsname{\color[rgb]{1,0.3,0}}%
      \expandafter\def\csname LT8\endcsname{\color[rgb]{0.5,0.5,0.5}}%
    \else
      \def\colorrgb#1{\color{black}}%
      \def\colorgray#1{\color[gray]{#1}}%
      \expandafter\def\csname LTw\endcsname{\color{white}}%
      \expandafter\def\csname LTb\endcsname{\color{black}}%
      \expandafter\def\csname LTa\endcsname{\color{black}}%
      \expandafter\def\csname LT0\endcsname{\color{black}}%
      \expandafter\def\csname LT1\endcsname{\color{black}}%
      \expandafter\def\csname LT2\endcsname{\color{black}}%
      \expandafter\def\csname LT3\endcsname{\color{black}}%
      \expandafter\def\csname LT4\endcsname{\color{black}}%
      \expandafter\def\csname LT5\endcsname{\color{black}}%
      \expandafter\def\csname LT6\endcsname{\color{black}}%
      \expandafter\def\csname LT7\endcsname{\color{black}}%
      \expandafter\def\csname LT8\endcsname{\color{black}}%
    \fi
  \fi
  \setlength{\unitlength}{0.0500bp}%
  \begin{picture}(5040.00,3528.00)%
    \gplgaddtomacro\gplbacktext{%
      \csname LTb\endcsname%
      \put(1188,897){\makebox(0,0)[r]{\strut{}0.0}}%
      \put(1188,1370){\makebox(0,0)[r]{\strut{}0.4}}%
      \put(1188,1844){\makebox(0,0)[r]{\strut{}0.8}}%
      \put(1188,2317){\makebox(0,0)[r]{\strut{}1.2}}%
      \put(1188,2791){\makebox(0,0)[r]{\strut{}1.6}}%
      \put(1188,3264){\makebox(0,0)[r]{\strut{}2.0}}%
      \put(1320,440){\makebox(0,0){\strut{} 0}}%
      \put(1878,440){\makebox(0,0){\strut{} 50}}%
      \put(2435,440){\makebox(0,0){\strut{} 100}}%
      \put(2993,440){\makebox(0,0){\strut{} 150}}%
      \put(3551,440){\makebox(0,0){\strut{} 200}}%
      \put(4108,440){\makebox(0,0){\strut{} 250}}%
      \put(4666,440){\makebox(0,0){\strut{} 300}}%
      \put(550,1962){\rotatebox{90}{\makebox(0,0){\strut{}$\delta \textrm{arg}\left [\Psi_4^{2,2}\right]$}}}%
      \put(2993,110){\makebox(0,0){\strut{}$t$}}%
    }%
    \gplgaddtomacro\gplfronttext{%
      \csname LTb\endcsname%
      \put(2307,3058){\makebox(0,0)[l]{\strut{}$a-b$}}%
      \csname LTb\endcsname%
      \put(2307,2772){\makebox(0,0)[l]{\strut{}$C_3 \times (b - f)$}}%
      \csname LTb\endcsname%
      \put(2307,2486){\makebox(0,0)[l]{\strut{}$C_4 \times (b - f)$}}%
      \csname LTb\endcsname%
      \put(2307,2200){\makebox(0,0)[l]{\strut{}$C_5 \times (b-f)$}}%
    }%
    \gplbacktext
    \put(0,0){\includegraphics{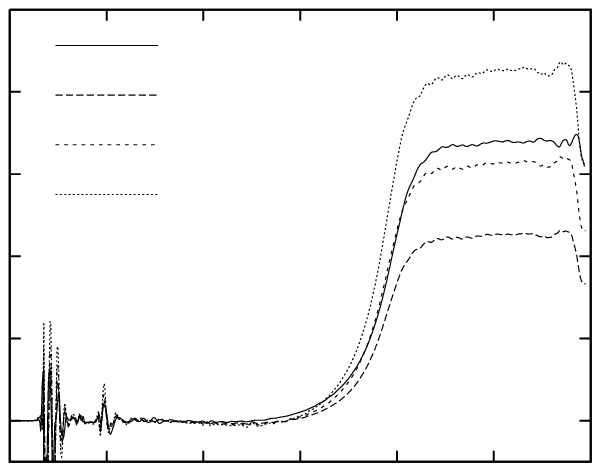}}%
    \gplfronttext
  \end{picture}%
\endgroup